\documentclass[12pt,preprint]{aastex} 

\def\gtorder{\mathrel{\raise.3ex\hbox{$>$}\mkern-14mu
             \lower0.6ex\hbox{$\sim$}}} 
\def\ltsima{$\; \buildrel < \over \sim \;$}
\def\simlt{\lower.5ex\hbox{\ltsima}}
\def\gtsima{$\; \buildrel > \over \sim \;$}
\def\simgt{\lower.5ex\hbox{\gtsima}} 

\begin{document} 


\title{Caltech Core-Collapse Project (CCCP) observations of type IIn supernovae:\\ 
typical properties and implications for their progenitor stars}


\author{Michael Kiewe, Avishay Gal-Yam, Iair Arcavi}
\affil{Benoziyo Center for Astrophysics, Faculty of Physics, The Weizmann
Institute of Science, Rehovot 76100, Israel}
\email{avishay.gal-yam@weizmann.ac.il}

\and

\author{Douglas C. Leonard, J. Emilio Enriquez}
\affil{Department of Astronomy, San Diego State University, San Diego, CA 92182.}

\author{S. Bradley Cenko}
\affil{Department of Astronomy, University of California, Berkeley, CA 94720-3411.}

\author{Derek B. Fox}
\affil{Department of Astronomy and Astrophysics, 525 Davey Laboratory, Pennsylvania State University, University Park, PA 16802.}

\author{Dae-Sik Moon}
\affil{Department of Astronomy and Astrophysics, University of Toronto, Toronto, ON M5S 3H4, Canada}

\author{David J. Sand, Alicia M. Soderberg}
\affil{Harvard-Smithsonian Center for Astrophysics, 60 Garden Street, Cambridge MA 02138}

\author{(the CCCP)}



\begin{abstract} 

Type IIn Supernovae (SNe IIn) are rare events, constituting only a few percent
of all core-collapse SNe, and the current sample of well observed
SNe IIn is small. Here, we study the four SNe IIn observed 
by the Caltech Core-Collapse Project (CCCP). The CCCP SN sample is unbiased 
to the extent that object selection was not influenced by target SN properties.
Therefore, these events are representative of the observed population of SNe IIn. 
We find that a narrow P-Cygni profile in the hydrogen Balmer lines appears
to be a ubiquitous feature of SNe IIn. Our light curves show a relatively long
rise time ($>20$ days) followed by a slow decline stage (0.01 to 0.15 mag day$^{-1}$),
and a typical $V$-band peak magnitude of $M_{V}=-18.4\pm1.0$ mag. We measure 
the progenitor star wind velocities ($600 - 1400$ km s$^{-1}$) for the 
SNe in our sample and derive pre-explosion mass loss 
rates ($0.026 - 0.12$ M$_{\odot}$ y$^{-1}$). We compile similar data 
for SNe IIn from the literature, and discuss our results in the 
context of this larger sample. 
Our results indicate that typical SNe IIn arise from progenitor stars
that undergo LBV-like mass-loss shortly before they explode. 

\end{abstract} 




\section{Introduction} 

For many decades it has been known that hydrogen-rich type II supernovae (SNe; 
Minkowski 1941; see Filippenko
1997 for a review on SN spectroscopic classification) arise from the 
gravitational core-collapse of massive stars (e.g., Arnett et al. 1989;
Smartt 2009). Schlegel (1990) recognized that 
a small subset of SNe II show narrow emission lines, and proposed that they constituted
a separate subclass (SNe IIn). It has been suggested that the narrow emission
lines of SN~IIn originate from photoionized dense wind surrounding the exploding
stars (Chugai 1991). 
It was later noticed that some SNe IIn (e.g., SN 1988Z) show, in addition to 
a narrow unresolved H$\alpha$ emission component ($\Delta v<200$\,km\,s$^{-1}$) 
also an intermediate component 
($\Delta v\sim10^3$\,km\,s$^{-1}$; Stathakis \& Sadler 1991). This component has been
interpreted as a result of radiative shocks in dense wind clouds
(Chugai \& Danziger 1994), though other explanations have been suggested
(Chugai 2001; Dessart et al. 2009).

Over the years, a specific model has emerged, which describes the properties
of SNe IIn as a result from a combination of emission from several physically distinct 
regions (see Smith et al. 2008 for a recent description). In brief, 
it is assumed that the SN ejecta interact with a massive
slower wind blown by the progenitor star, launching a pair of forward and
reverse shocks. The blast wave is decelerated by the interaction from
its initial velocity, and a dense shell traveling approximately at
the shock velocity is formed. Initially, this dense shell is optically
thick, and obscures the actual electron-scattering photosphere of the supernova which lies interior to the shock.
At this stage (which can, in principle, have a duration of years),
the optical spectrum is dominated by a blue continuum from the shocked
material, superposed with emission lines. The emission lines often have
a complex profile, including a narrow component from the photoionized, 
outer unshocked wind, and an intermediate-width component from the
shocked dense shell. Later on, as the interaction weakens, the 
shocked region may become optically thin and the underlying supernova
photosphere (with its high-velocity absorption-dominated spectrum) may
emerge (e.g., Gal-Yam \& Leonard 2009). Applying such a model, one can extract the physical
properties of the progenitor star (in particular, its mass-loss rate)
from the observations (e.g., Salamanca et al. 1998, Gal-Yam \& Leonard 2009). 
Previously derived values (summarized in Table~\ref{oldSNIIntable}) show progenitor
wind velocities of $50-1000$\,km\,s$^{-1}$ and mass loss rates that range between 
$10^{-4} - 0.3$\,M$_{\odot}$\,y$^{-1}$. These are consistent with a variety of 
possible massive progenitors, from red supergiant stars (RSGs;
typical wind velocities of tens of km\,s$^{-1}$, e.g., Smith et al. 2007) and small mass-loss rates
($\simlt10^{-5}$\,M$_{\odot}$\,y$^{-1}$), to stripped Wolf-Rayet (W-R) stars with fast
(thousands of km\,s$^{-1}$) and more massive winds ($\sim10^{-4}$\,M$_{\odot}$\,y$^{-1}$), 
up to Luminous Blue Variable stars (LBVs) which have the most extreme mass-loss 
rates (up to $\sim10^{-2}$\,M$_{\odot}$\,y$^{-1}$) and intermediate-velocity
winds (a few hundreds of km\,s$^{-1}$; e.g., Humphreys \& Davidson 1994).  

The connection between SNe IIn and LBVs was brought into focus by studies
of SN 2005gl. Gal-Yam et al. (2007a) analyzed pre-explosion images of this
SN IIn, and detected a possible progenitor, whose extreme luminosity 
(L$_V>10^{6}$\,L$_{\odot}$) suggested an LBV identification (the only 
stars of comparable properties known in our galaxy are LBVs). 
Gal-Yam et al. (2007) also noted that the relative rarity and spectral appearance of SNe IIn could be naturally explained by an origin in very 
massive LBVs (with M$\simgt80$\,M$_{\odot}$). Gal-Yam \& Leonard (2009) subsequently
presented post-explosion HST data of this event that showed the putative 
progenitor of SN 2005gl had disappeared, providing strong evidence 
that this SN IIn had indeed originated from a single massive LBV-analog. 
These authors also analyzed the spectra of SN 2005gl and showed that the derived mass-loss rate
was consistent with that of eruptive LBVs (e.g., P-Cygni during its sixteenth
century eruption). Other work on additional SNe supports the SN IIn-LBV
connection (e.g., Smith et al. 2007; Trundle et al. 2008).

It remains an open question how general the SN IIn-LBV connection is, i.e., are 
most or all SNe IIn related to LBVs, or stars with similar properties? 
A literature-based study of this question is weakened due to the fact that 
the sample of SNe IIn published thus far may be highly biased, due to the tendency 
of authors to preferentially publish extreme or unusual events (this is 
evident for many of the objects, as summarized in section $\S$~\ref{historicSNeIIn}
and Table~\ref{oldSNIIntable}). 

Here we present and analyze four SNe IIn observed by the Caltech Core-Collapse 
Project (CCCP; Gal-Yam et al. 2007b; Gal-Yam et al. 2010, in preparation). 
The CCCP supernova sample is unbiased in the sense that essentially 
every young core-collapse SN that was observable from Palomar Observatory during the project
was followed-up, without preference. Thus, the objects included within the CCCP
should fairly represent the population of observed SNe IIn. Typical wind velocities
and mass-loss rates can therefore be derived, which may help constrain the properties of the progenitor stars. We organize the paper 
as follows. In $\S~2$ we present our observations, and derive physical parameters for the sample 
in $\S~3$. We present an exhaustive literature review of SNe IIn in $\S~4$, and discuss
our results in $\S~5$. We conclude with a summary in $\S~6$.

\section{Observations}

The observations reported here were taken as part of the CCCP (Gal-Yam et al. 2007b, Gal-Yam et al. 2010 in preparation). 
The CCCP obtained photometric (optical and near-IR) and spectroscopic
observations of 50 core-collapse SNe. The program is
designed to provide a fair sample of core-collapse events, with well-defined
selection criteria and uniform, high-quality optical/IR observations.
One of the goals of the CCCP is to characterize the little-studied
properties of core-collapse SNe as a population. 4 SNe IIn are included
in the CCCP SN sample. 

\subsection{Discovery}
\label{discoverysection}

SN 2005bx was discovered by Rich (2005) on April 27, 2005,
and was absent on April 11, 2005. It was thus discovered within 16 days of explosion. The host
galaxy, MCG +12-13-19 is at z=0.031218 (NED\footnote{see http://nedwww.ipac.caltech.edu}). 
Bonnaud et al. (2005) identified SN 2005bx as an SN IIn based
on a spectrum taken on May 1, 2005; our earlier spectrum from April 30 (Gal-Yam et al. 2005) confirms this identification. 

Pugh \& Li (2005) reported the discovery of SN 2005cl 
by LOSS (Filippenko et al. 2001). The SN was detected on June 2 (marginal) 
and June 12, 2005, and was absent on May 23, 2005,
so it too was discovered rather young. The host galaxy MCG -01-53-20 is at z=0.025878
(NED). A spectrum taken by Modjaz et al. (2005a) on June 13, 2005 demonstrates
that SN 2005cl was a SN IIn.

Lee et al. (2005) reported the LOSS (Filippenko et al. 2001) discovery of 
SN 2005cp on June 20, 2005.  Spectroscopy by Modjaz et al.
(2005b) on June 30, 2005 indicated a young age and IIn classification. The host
galaxy, UGC 12886, is at z=0.022115 (NED). Our light curve confirms
that this was a young SN found on the rise (see Fig.~\ref{figlc05cp} below).

SN 2005db was discovered by Monard (2005) on July 19, 2005, and was
absent on July 2, 2005, so it was discovered rather early. The host galaxy,
NGC 214, is at z=0.015124 (NED). Blanc et al. (2005) identified this
event as a type IIn based on a spectrum taken on July 20, 2005. Our light curve confirms that this was
a young SN found on the rise (Fig.~\ref{figlc05db}).

\subsection{Photometry}
\label{photsection}

We obtained optical photometry with the 1.5\,m robotic telescope at
the Palomar Observatory (P60; Cenko et al. 2006) using a 2048 x 2048 pixel CCD
in the Johnson-Cousins $BVRI$ bands. All images were pipeline-reduced by the
automated P60 software, applying standard bias- and flatfield-correction,
as well as an astrometric solution.  

We derived light curves using the image-subtraction-based photometry 
routine mkdifflc (Gal-Yam et al. 2004), implemented within IRAF\footnote{IRAF 
(Image Reduction and Analysis Facility) is distributed by the
National Optical Astronomy Observatories, which are operated by
AURA, Inc., under cooperative agreement with the National Science
Foundation.}. This pipeline performs image registration followed by 
reference image subtraction using the Common PSF Method (CPM; Gal-Yam et al. 
2008), relative calibration with respect to local standards, and error
calculation.

Field calibration was carried out relative to stars near the SN location
with absolute photometry obtained in two ways. In the fields of SN 2005cl and SN
2005db, SDSS cataloged $gri$ magnitudes of nearby stars were transformed
to the $BVRI$ Vega-based system using the
equations of Jordi et al (2005). The fields of SN 2005bx and SN 2005cp
are not within the SDSS footprint. For those we obtained standard
photometric calibration using Landolt standards taken by the P60 on photometric
nights. Additional discussion of photometric calibration, as well as
finding charts for the nearby calibration stars used for the 4 SN fields 
in this work, are given in the appendix. 

Photometry of all SNe was corrected for Galactic extinction (Schlegel, Finkbeiner
\& Davis 1998; via NED). The final light curves were also corrected for extinction
in the host galaxy by first estimating using the IRAF {\it splot} routine 
the equivalent width (EW) of the Na I D absorption feature in our spectra 
and then calculating the excess color E(B-V) from the EW using the formula
of Turatto et al. (2003). In two cases (SN 2005cl and SN 2005bx) an apparently 
unrelated noise fluctuation falls close to the Na D feature in our best spectra.
We also calculate an upper limit on the extinction assuming this apparent noise feature
is due entirely to Na D absorption. We view this upper limit as very conservative.
Therefore, we corrected the light curves shown for the most likely estimate reported in the 
table. In the mass loss calculations below we adopt the average between the likely
estimate and upper limit as our extinction value, and propagate the range bracketed 
between these values in the mass loss error calculations.  
The adopted extinctions are given in Table~\ref{exttable}.

\begin{table}
\begin{tabular}{lcccc}
\hline 
\hline
                              & SN 2005bx & SN 2005cl & SN 2005cp & SN 2005db \tabularnewline
\hline 
E(B-V)$_{Galactic}$ [mag]     &     0.019 &     0.073 &     0.031 &     0.036 \tabularnewline 
\hline
Na I D EW (likely)[Ang]       &      0.67 &      0.73 &      0.12 &      0.58 \tabularnewline 
E(B-V)$_{Host}$ [mag]         &       0.3 &      0.33 &      0.02 &      0.26 \tabularnewline 
\hline
Na I D EW (upper limit)[Ang]  &      1.09 &      0.95 &           &           \tabularnewline 
E(B-V)$_{Host,limit}$ [mag]   &      0.52 &      0.45 &           &           \tabularnewline 
\hline
\hline
\end{tabular}
\caption{Adopted extinction values}
\label{exttable}
\end{table}

Late images (mid-end November 2005) of SN 2005db do not show signal from the
SN. An upper limit on its luminosity was therefore calculated by introducing artificial
point sources of decreasing flux at the SN location and using the SEXTRACTOR 
software (Bertin \& Arnouts 1996) to determine when the artificial object was detected at 3$\sigma$ 
confidence. These values are adopted as the reported upper limits.

\begin{figure}
\includegraphics[width=1\textwidth]{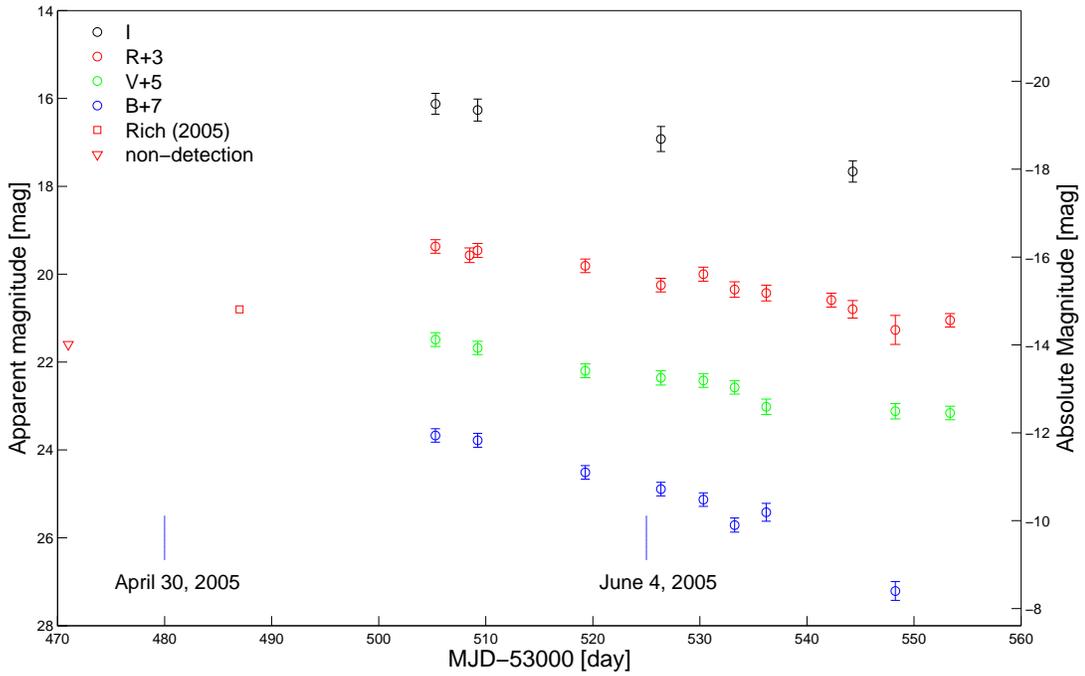}
\caption{$BVRI$ light curves of SN 2005bx. Discovery magnitude and
previous non-detection from Rich (2005) are plotted, on an $R$-band grid. 
Our follow-up campaign missed the peak (values given in table~\ref{lcparamtable}
are for our first, brightest point). The $VRI$ bands have a similar
decline rate, while the $B$-band declines approximately
twice as fast. 
The dashed blue lines mark the spectroscopic observation epochs of
SN 2005bx.}
\label{figlc05bx}
\end{figure}

\begin{figure}
\includegraphics[width=1\textwidth]{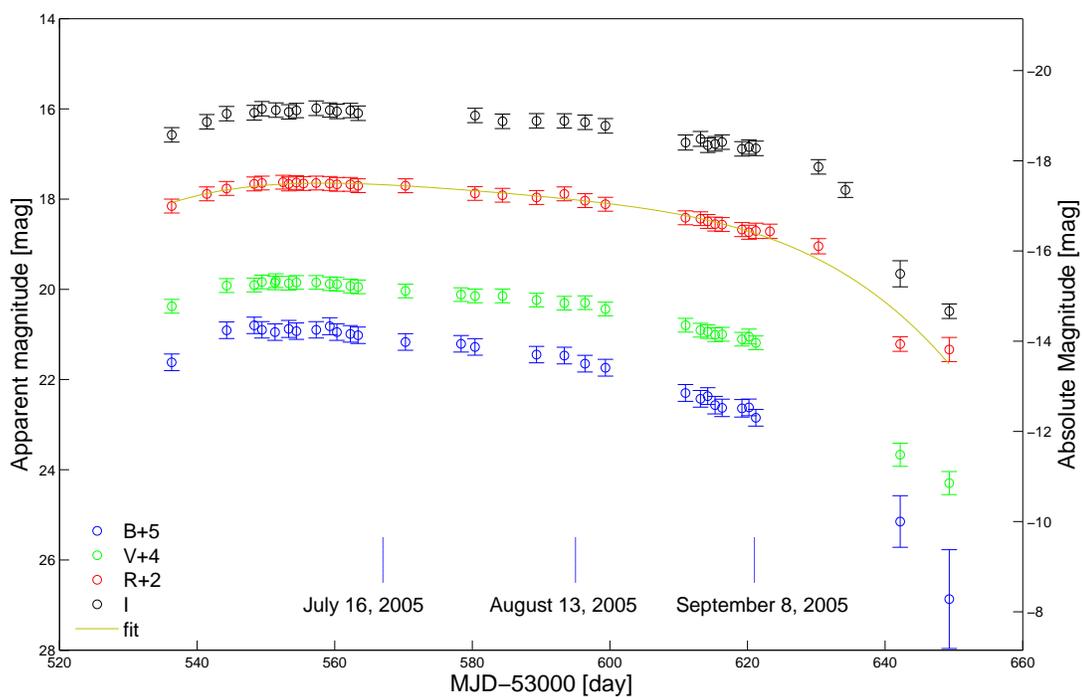}
\caption{$BVRI$ light curves of SN 2005cl. Our data trace a slow rise 
to peak magnitude, followed by a slow decay, with an obvious break
to a steeper decline around MJD 53625 day. The
blue vertical lines mark the spectroscopic observation epochs of SN 2005cl.}
\label{figlc05cl}
\end{figure}

\begin{figure}
\includegraphics[width=1\textwidth]{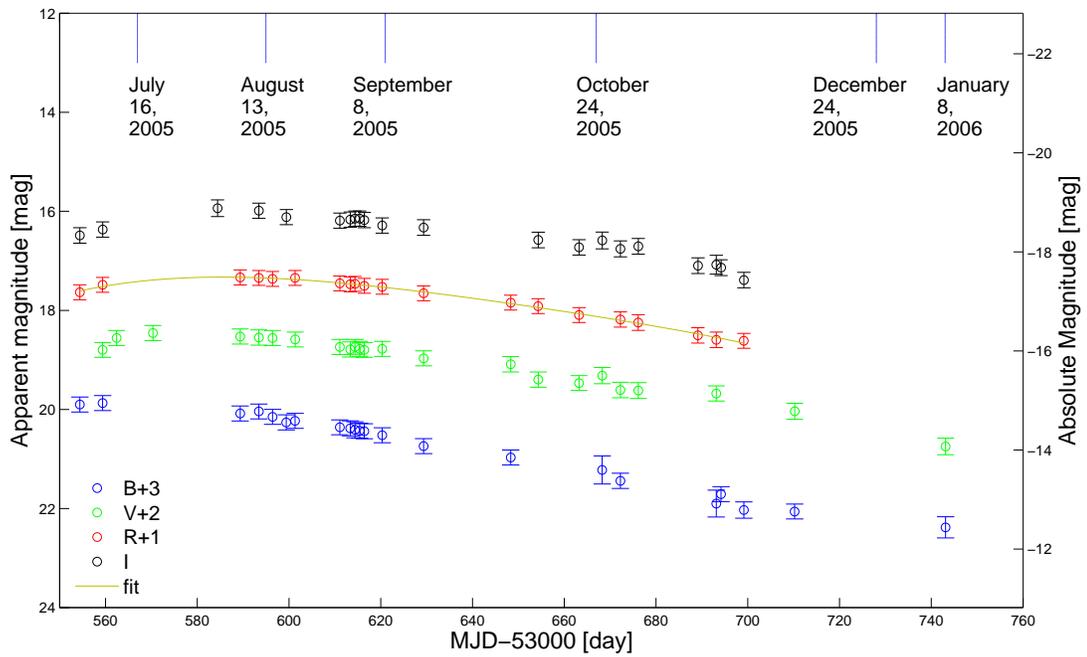}
\caption{$BVRI$ light curves of SN 2005cp. The SN exhibits a slow rise to peak,
followed by an extended decline stage of constant slope. The
blue vertical lines mark the spectroscopic observation epochs of SN
2005cp.}
\label{figlc05cp}
\end{figure}

\begin{figure}
\includegraphics[width=1\textwidth]{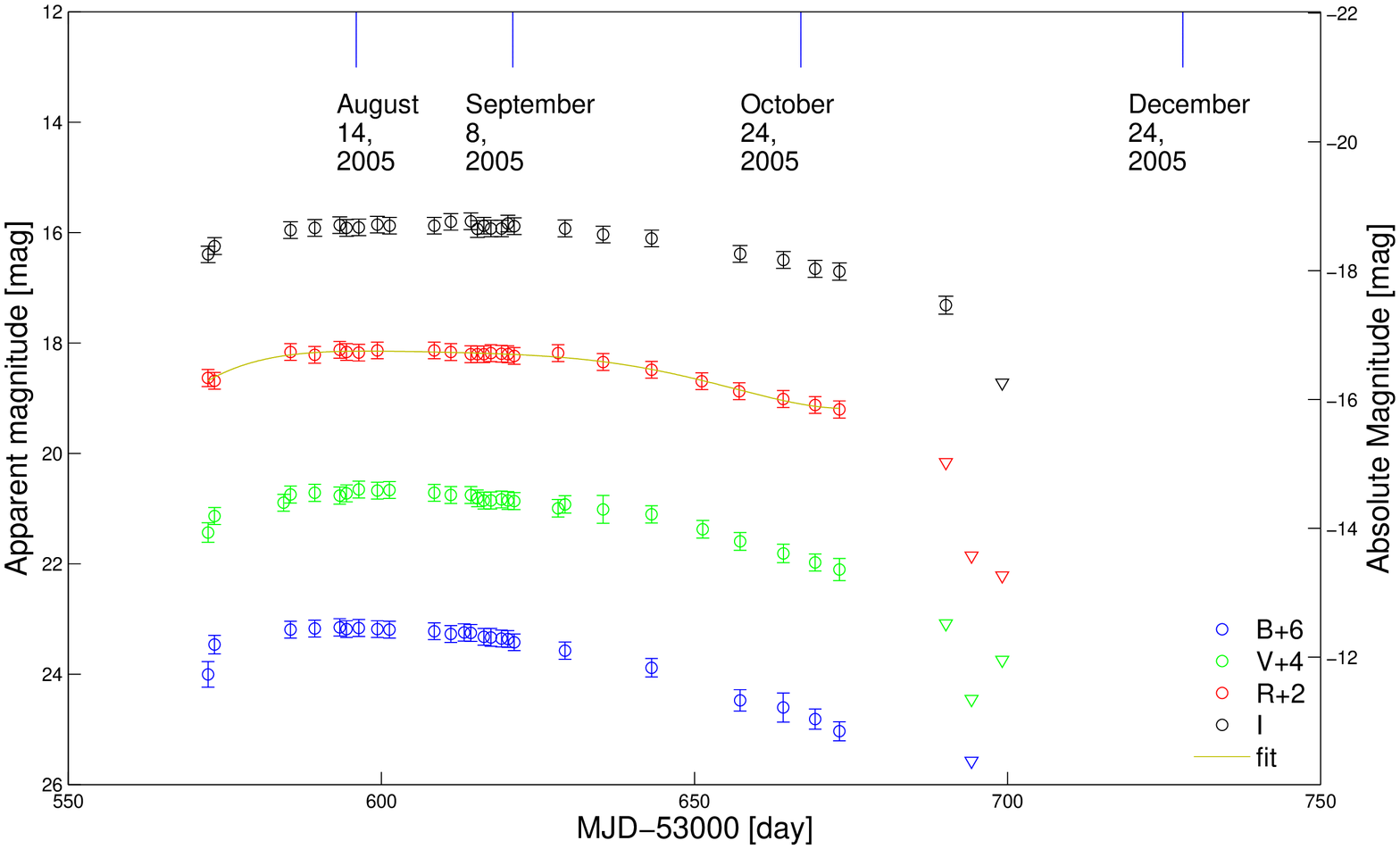}
\caption{$BVRI$ light curves of SN 2005db. The SN has an initially
rapid rise followed by a broad peak. Following the peak, there is
a period of 30-40 days in which the SN magnitude is almost constant. 
This is followed by a slow decline which is faster in bluer bands,
that eventually turns into a sharp
drop in the SN luminosity, traced by the upper limits marked
by downward-facing triangles. This magnitude drop can be evaluated in the 
$VRI$ bands to be between 1-2 mag within 4-9 days. This sharp drop in magnitude may be
indicative of a CSM distribution with a sharp cutoff, leading to an
abrupt luminosity decrease as the ejecta emerge from 
the previously ejected mass shell and the interaction effectively stops. The dashed
blue lines mark the spectroscopic observation epochs of SN 2005db.}
\label{figlc05db}
\end{figure}

The photometric light curves of all 4 SNe 
are presented in Figures~\ref{figlc05bx} -- \ref{figlc05db}. 
The errors include photometric errors due to non-perfect image subtraction and Poisson errors (derived by {\it mkdifflc}), extinction uncertainty 
errors, zeropoint calibration errors, and distance modulus errors taken
from NED, which usually dominate the errors in peak absolute magnitudes ($\approx0.15$\,mag). We determined peak
magnitudes and dates from low-order polynomial fits to the data (example fits
to the $R$-band data are presented). Rise times were estimated using our
data combined with reported non-detections, assumed to have been obtained in 
bands closely matching the $R$-band (section~\ref{discoverysection}).
Post-peak decay slopes were determined
from linear fits. SN 2005cl showed a clear break in the decline rate, so 
a broken linear fit was used. The BVRI peak values,
dates, rise times and decline rates are listed in table~\ref{lcparamtable}.

\begin{table}[h]
\begin{tabular}{cccccc}
\hline 
SN 2005bx &  &  &  & \tabularnewline
\hline 
Band & Peak magnitude & MJD & Rise time & Initial slope & Final slope \\
     & [mag]          & [day] & [day]   & [mag/day]     & [mag/day]   \\
\hline 
B    & $-18.94\pm0.15$      & 53505   &                  & $0.077\pm0.019$  &            \\
V    & $-19.12\pm0.15$      & 53505   &                  & $0.036\pm0.008$  &            \\
R    & $-19.24\pm0.15$      & 53505   & $<34$            & $0.037\pm0.006$  &            \\
I    & $-19.49\pm0.15$      & 53505   &                  & $0.040\pm0.002$  &            \\
\hline 
SN 2005cl &  &  &  & \tabularnewline
\hline 
B    & $-19.28\pm0.15$      & 53553   &                  & $0.026\pm0.003$  & $0.14\pm0.06$  \\
V    & $-19.32\pm0.15$      & 53554   &                  & $0.018\pm0.002$  & $0.11\pm0.07$  \\
R    & $-19.50\pm0.15$      & 53556   & $14-34$          & $0.016\pm0.002$  & $0.11\pm0.09$  \\
I    & $-19.14\pm0.15$      & 53557   &                  & $0.012\pm0.002$  & $0.18\pm0.07$  \\
\hline 
SN 2005cp &  &  &  & \tabularnewline
\hline 
B    & $-17.98\pm0.15$      & 53565   &                  & $0.016\pm0.001$  &            \\
V    & $-18.17\pm0.15$      & 53581   &                  & $0.014\pm0.001$  &            \\
R    & $-18.48\pm0.15$      & 53584   & $>32$            & $0.012\pm0.001$  &            \\
I    & $-18.85\pm0.15$      & 53586   &                  & $0.011\pm0.001$  &            \\
\hline 
SN 2005db &  &  &  & \tabularnewline
\hline 
B    & $-16.89\pm0.15$      & 53587   &                  & $0.028\pm0.002$  &            \\
V    & $-17.32\pm0.15$      & 53598   &                  & $0.020\pm0.002$  &            \\
R    & $-17.87\pm0.15$      & 53597   & $10-27$          & $0.011\pm0.001$  &            \\
I    & $-18.17\pm0.15$      & 53607   &                  & $0.021\pm0.004$  &            \\
\hline 
\end{tabular}
\caption{Photometric light curve parameters for SNe 2005bx, 2005cl, 2005cp, 2005db.}
\label{lcparamtable}
\end{table}

\clearpage

\subsection{Spectroscopy}

Spectroscopic observations
were carried out using the double beam spectrograph (DBSP; Oke \& Gunn 1982) 
mounted on the 5\,m Hale telescope at the Palomar Observatory. We used the 
D55 dichroic, with 600/158 line/mm grisms in the blue/red sides, respectively.
We obtained a continuous spectrum covering
the wavelength range $3400-9400$ \AA. The instrumental
resolution was $4.88$ \AA\ per pixel in the red and $1.08$ \AA\ per pixel
in the blue. The typical exposure time was between 10 and 30
minutes, see table \ref{obslogtable} for more details.

Spectroscopic reduction was performed using the CCCP spectroscopic 
pipeline, which is based on IRAF and IDL scripts as described, e.g., in 
Matheson et al. (2000), and Gal-Yam et al. (2007). 
In brief, to derive the total-flux spectrum, 
we extracted the one dimensional sky-subtracted spectra optimally 
(Horne 1986) in the usual manner. The spectra were then wavelength 
and flux calibrated, corrected for continuum atmospheric extinction 
and telluric absorption bands (Wade \& Horne 1988; Bessell 1999; Matheson
et al. 2000).  

\begin{table}
\begin{tabular}{ccccc}
\hline 
Supernova & UT Date & exposure time [sec] & Observers & Comments \tabularnewline
\hline
\hline 
2005bx & Apr. 30, 2005 & 900  & Sand \& Cenko     & - \tabularnewline 
       & June 4, 2005  & 1200 & Gal-Yam \& Sand   & - \tabularnewline
       & July 16, 2005 & 1200 & Gal-Yam \& Sharon & slightly cloudy \tabularnewline
\hline 
2005cl & July 16, 2005 & 600  & Gal-Yam \& Sharon & - \tabularnewline 
       & Aug. 13, 2005 & 1800 & Cenko             & - \tabularnewline 
       & Sep. 8, 2005  & 1800 & Leonard           & clear sky, poor seeing \tabularnewline
\hline 
2005cp & July 16, 2005 & 600  & Gal-Yam \& Sharon & clear sky \tabularnewline
       & Aug. 13, 2005 & 1800 & Cenko             & - \tabularnewline 
       & Sep. 8, 2005  & 1800 & Leonard           & clear sky, poor seeing \tabularnewline 
       & Oct. 24, 2005 & 1800 & Gal-Yam           & - \tabularnewline
       & Dec. 24, 2005 & 1800 & Sand              & - \tabularnewline
       & Jan. 8, 2006  & 1800 & Cenko \& Ballmer  & - \tabularnewline
       & Jan. 20, 2006 & 1800 & Moon              & - \tabularnewline
\hline 
2005db & Aug. 14, 2005 & 1200 & Sand              & - \tabularnewline 
       & Sep. 8, 2005  & 900  & Leonard           & - \tabularnewline
       & Oct. 24, 2005 & 900  & Gal-Yam           & clear \tabularnewline
       & Dec. 24, 2006 & 1800 & Sand              & poor seeing \tabularnewline
       & Jan. 8, 2006  & 1800 & Cenko \& Ballmer  & - \tabularnewline
       & Jan. 20, 2006 & 1800 & Moon              & -\tabularnewline
\hline
\hline
\end{tabular}
\caption{Log of spectroscopic observations}
\label{obslogtable}
\end{table}

Our resulting spectroscopic time series are presented in 
figures~\ref{specfig05bx}-\ref{specfig05db}. We discuss each
object separately below. 

\clearpage

\subsubsection{SN 2005bx}

\begin{figure}
\includegraphics[width=1\textwidth]{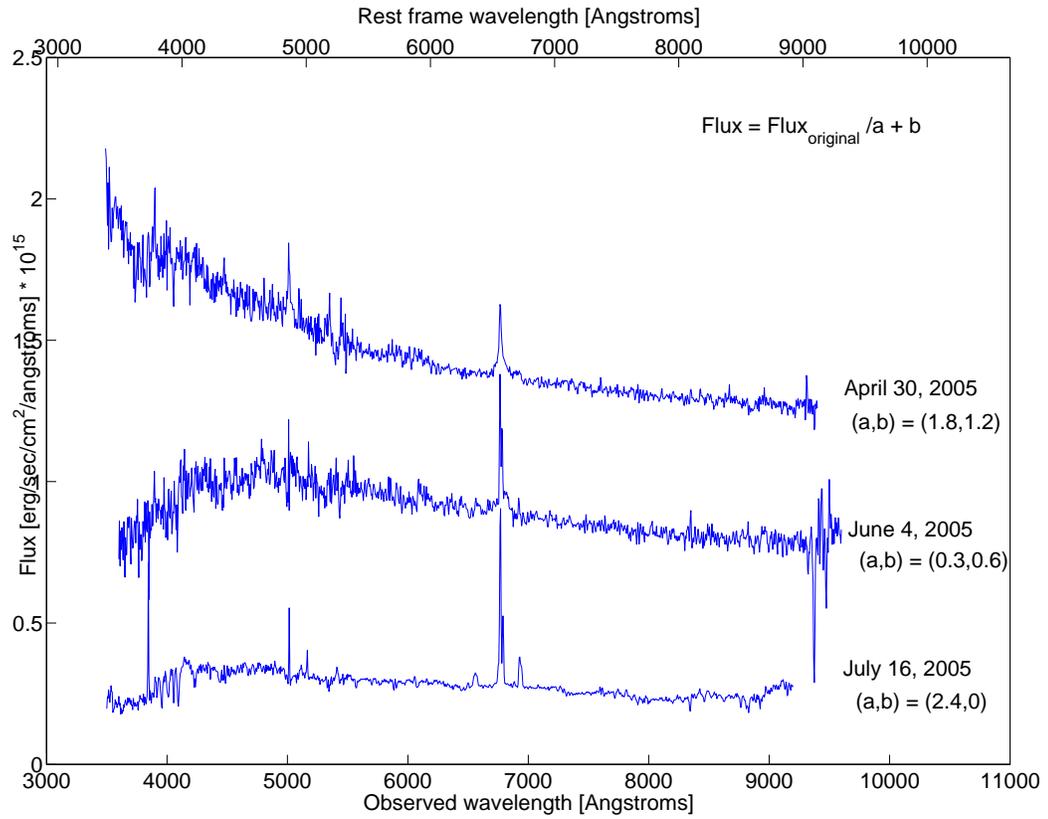}
\caption{Spectral evolution of SN2005bx. The intermediate
component evident in the first spectrum fades away in later observations.}
\label{specfig05bx}
\end{figure}

We present our spectroscopy of SN2005bx in Fig.~\ref{specfig05bx}. 
The continuum shape was initially blue, and reddened with time. Except
for the Balmer emission lines, there are no other strong features. 
The intermediate-width component of the H$\alpha$ line is evident in 
the first spectrum, which we decompose in Fig.~\ref{decomp2005bx}
below, and disappears in later spectra. None of our spectra exhibit
P-Cygni profiles in the narrow lines.

\clearpage

\subsubsection{SN 2005cl}

\begin{figure}
\includegraphics[width=1\textwidth]{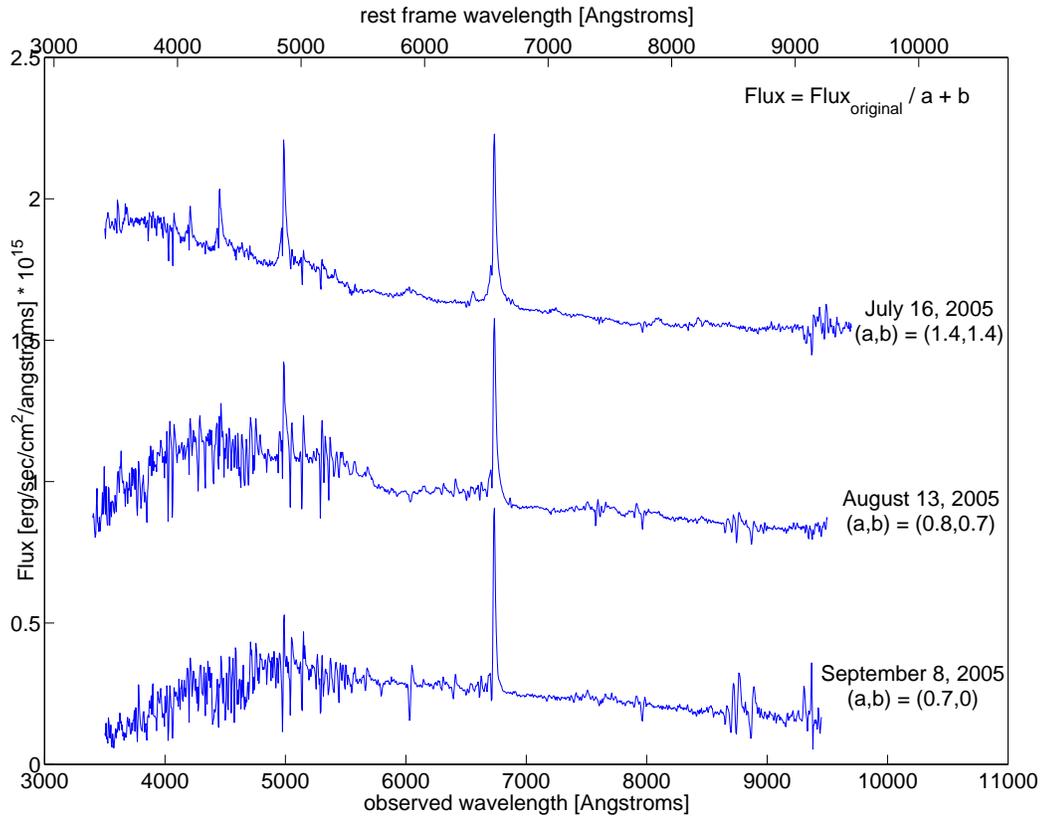}
\caption{Spectral evolution of SN2005cl. One can see that the intermediate
component of the Balmer lines fades away in the later observations.
Note the narrow P-Cygni profile in the 4 first Balmer lines in the
earliest spectrum.}
\label{specfig05cl}
\end{figure}

\begin{figure}
\includegraphics[width=1\textwidth]{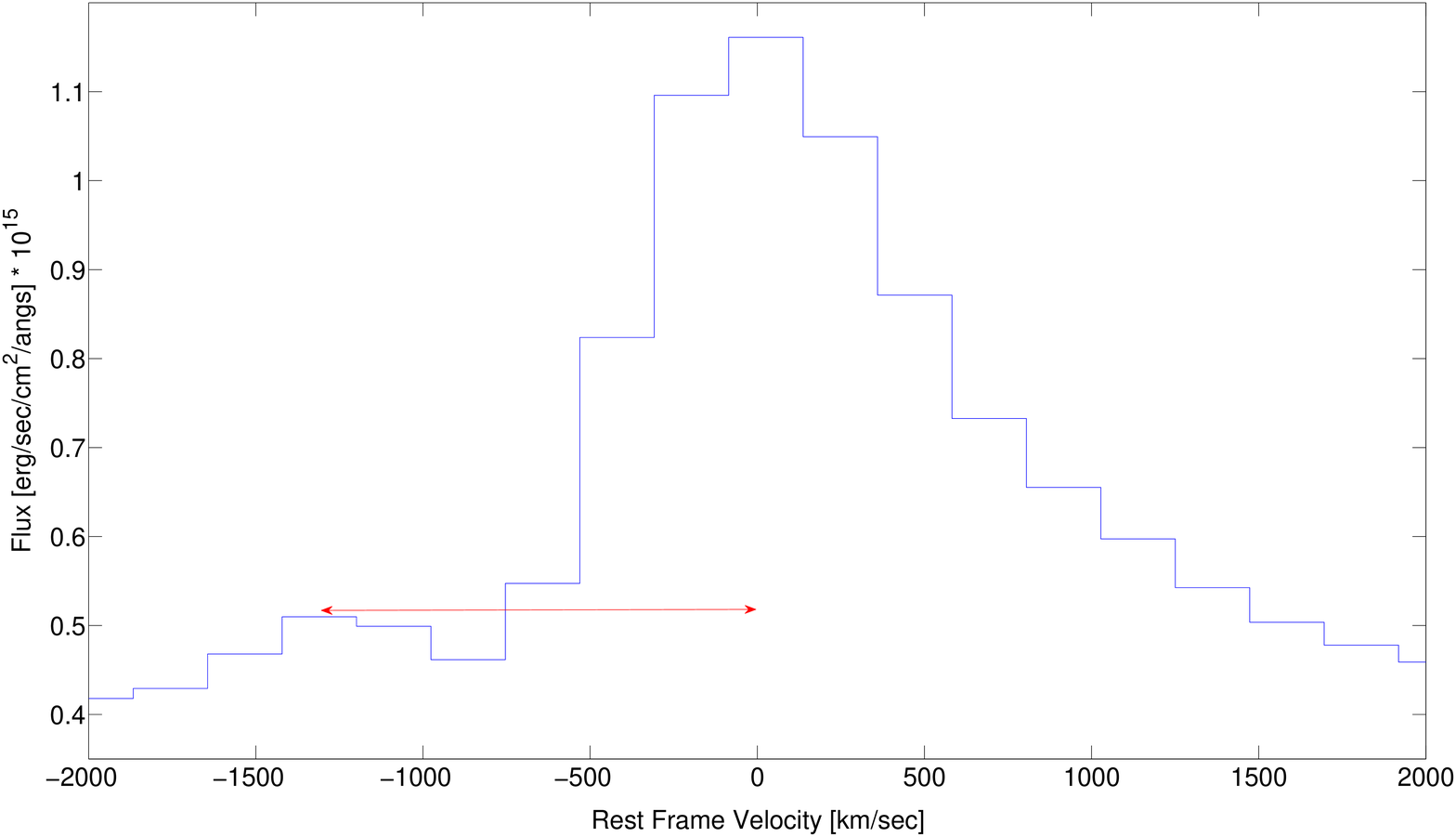}
\caption{The H$_{\alpha}$ emission line in our first spectrum of SN2005cl 
shows a clear P-Cygni profile. The red arrow marks the velocity offset of the blue edge
of the absorption feature, used to estimate the progenitor wind velocity.}
\label{sn2005cl-halpha-zoom}
\end{figure}

\begin{figure}
\includegraphics[width=1\textwidth]{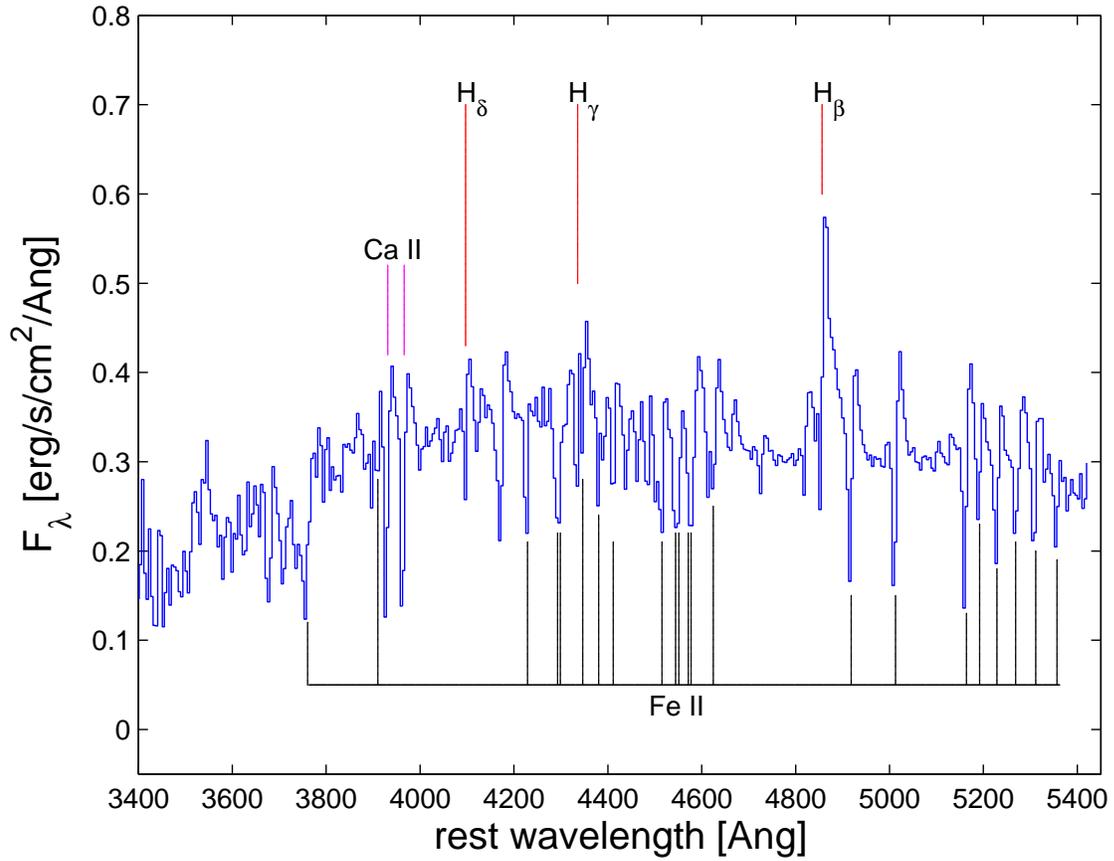}
\caption{A zoomed-in section of the Sep. 8, 2005 spectrum of SN 2005cl. Note
that the blue continuum bump is partially resolved into multiple narrow P-Cygni
profiles of H, Fe II and Ca II.}
\label{pcont}
\end{figure}

We present our spectroscopy of SN2005cl in Fig.~\ref{specfig05cl}, showing
the spectral evolution of the SN. One can see the hallmark of a
type IIn SN in the form of narrow hydrogen emission lines with a P-Cygni
profile (that indicating the presence of an expanding envelope) in
$H_{\alpha}$, $H_{\beta}$, $H_{\gamma}$ and $H_{\delta}$. The
unshocked wind velocity was calculated from the $H_{\alpha}$ 
P-Cygni profile (Fig.~\ref{sn2005cl-halpha-zoom}). 
In Fig.~\ref{decomp2005cl} we present the decomposition
of the $H_{\alpha}$ feature in our first spectrum into different 
Gaussian components, showing the presence of an intermediate-width
component. In later spectra this
component is not clearly visible, while the narrow P-Cygni profiles 
in the Balmer lines persist. The relatively high resolution of our blue spectra
reveals that in these later epochs, the continuum ``bump'' on the blue
is actually a pseudo-continuum, and is at least partially resolved into 
numerous very narrow P-Cygni profiles, mainly of Fe II lines (Fig.~\ref{pcont}). 

\clearpage

\subsubsection{SN 2005cp}

We present our spectroscopy of SN2005cp in Fig.~\ref{specfig05cl}. 
The first spectrum shows a blue continuum bump (not resolved into 
individual lines) and prominent Balmer emission lines. In later spectra
the blue continuum fades, while SN photospheric features emerge, including
a broad base below the strong Balmer emission lines, a broad line with
P-Cygni profile blueward of $6000$\AA\ which could be He I $5876$\AA\
(or Na D), and later on, a broad Ca IR triplet. A strong $H_{\alpha}$ emission
dominates the spectrum at very late times.   
In Fig.~\ref{decomp2005cp}, we zoom
in on the $H_{\alpha}$ feature in the August spectrum and extract
the unshocked wind velocity from the decomposition of the $H_{\alpha}$
feature into different Gaussian components. The July and September
spectra were analyzed in the same way and provided consistent results.
In Fig.~\ref{sn2005cp-halpha-zoom}, we zoom in on the $H_{\alpha}$ feature in the January
8th spectrum and extract the unshocked wind velocity from the P-Cygni
profile.

\begin{figure}
\includegraphics[width=1\textwidth]{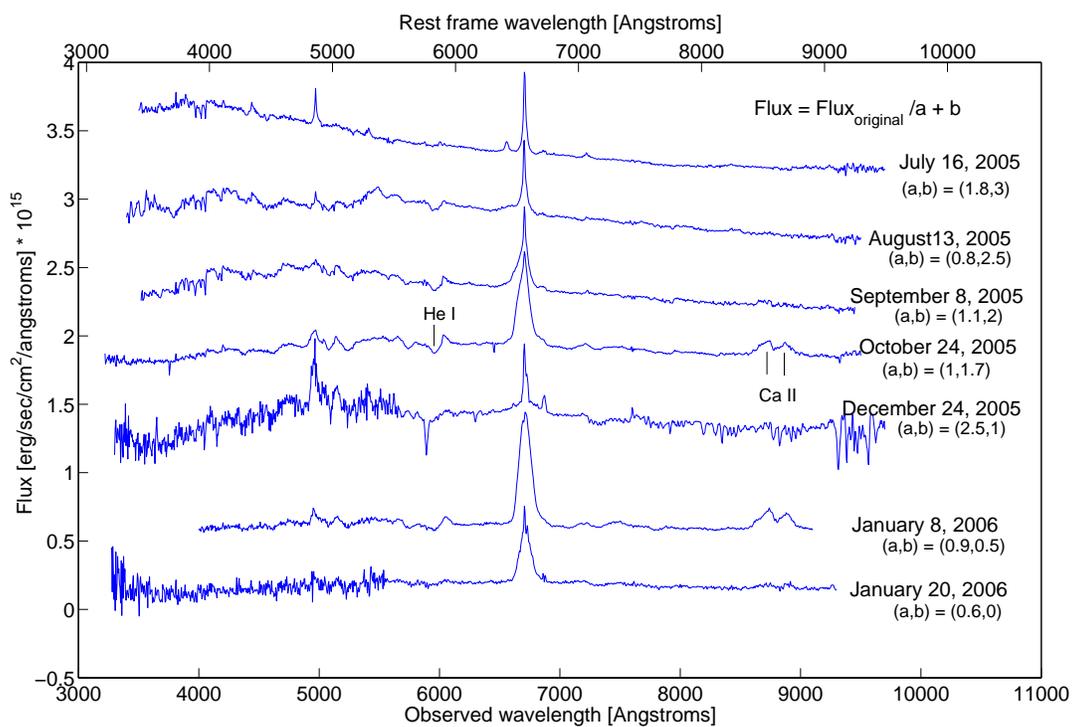}
\caption{Spectral evolution of SN2005cp. Broad, intermediate, and narrow
components are prominent in the August 13 spectrum (second from top). 
The blue continuum in early spectra is not resolved into individual lines,
and narrow P-Cygni features appear only in the $H_{\alpha}$ feature in a late
spectrum. A strong $H_{\alpha}$ emission line dominates the spectrum at 
late times.}
\label{specfig05cp}
\end{figure}

\begin{figure}
\includegraphics[width=1\textwidth]{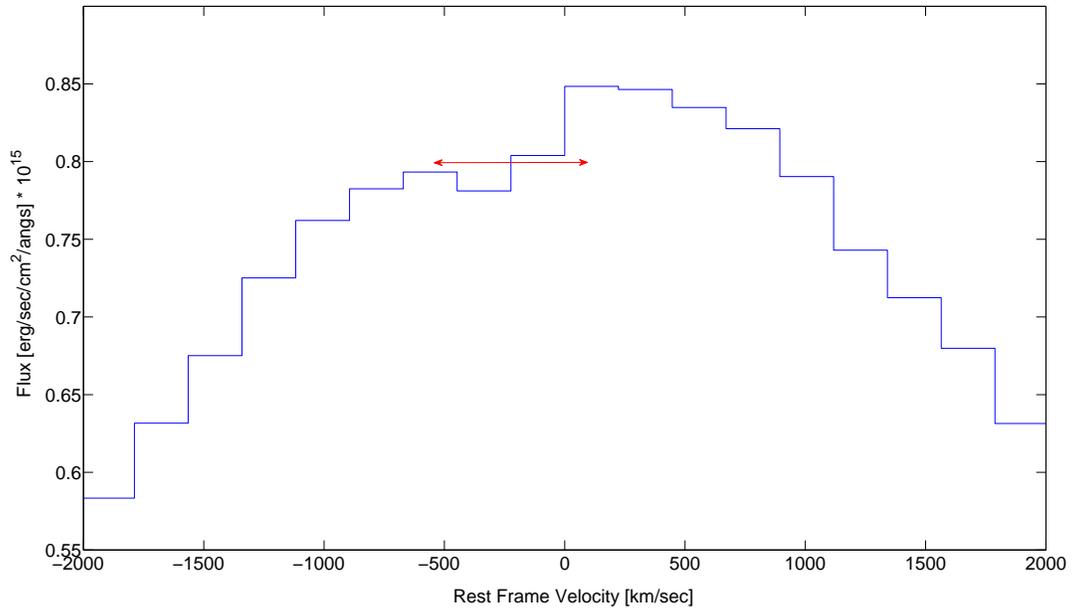}
\caption{The H$_{\alpha}$ emission line in our January 8 spectrum of SN2005cp 
shows a clear P-Cygni profile. The red arrow marks the velocity offset of the blue edge
of the absorption feature, used to estimate the progenitor wind velocity.}
\label{sn2005cp-halpha-zoom}
\end{figure}

\clearpage

\subsubsection{SN 2005db}

\begin{figure}
\includegraphics[width=1\textwidth]{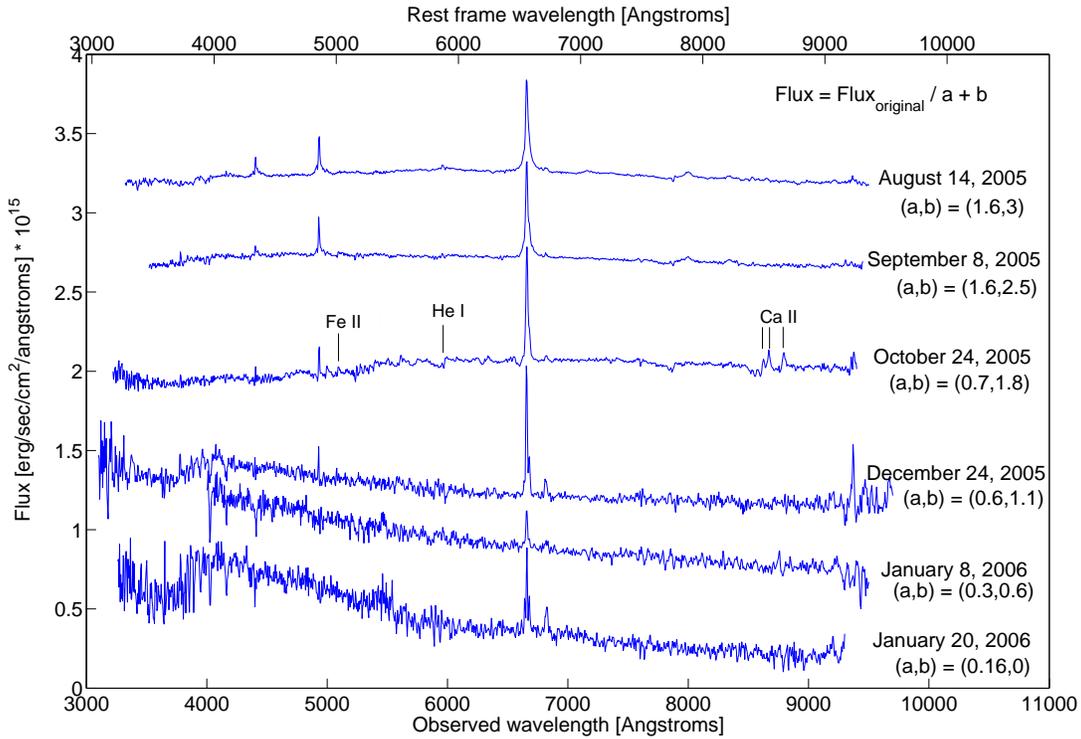}
\caption{Spectral evolution of SN2005db. Composite profiles of the 
Balmer lines are evident, with narrow P-Cygni profiles. The October spectrum
shows emerging intermediate-width P-Cygni profiles of Fe II, He I and 
the Ca II IR triplet. 
The last 3 spectra appear to be dominated
by the host galaxy light, similar to a starburst template spectrum
from Kinney (1996).}
\label{specfig05db}
\end{figure}

\begin{figure}
\includegraphics[width=1\textwidth]{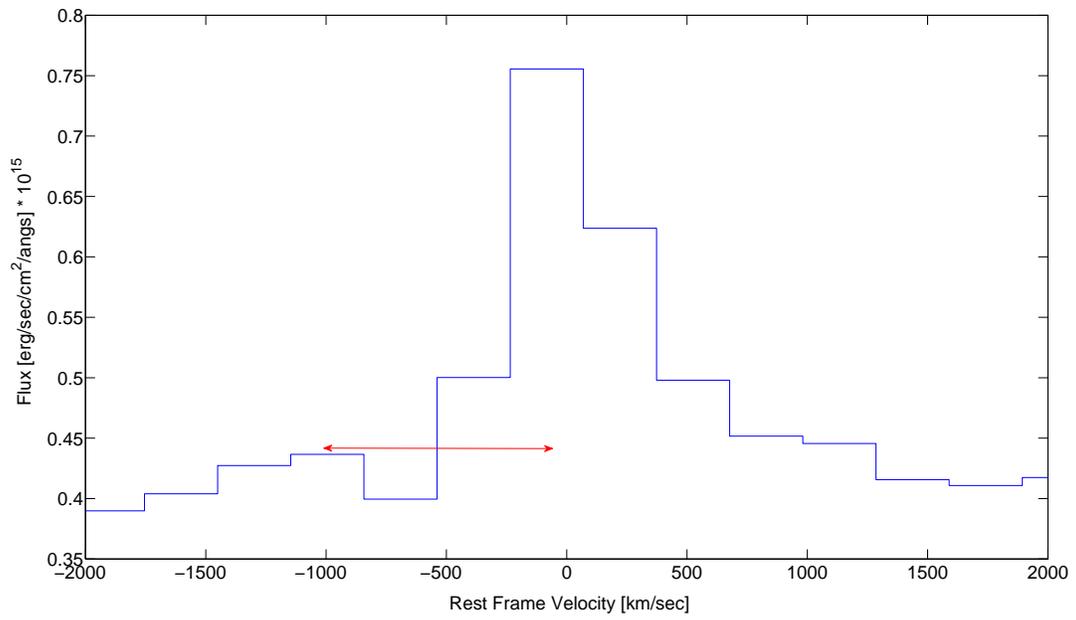}
\caption{A Zoom-in on the $H_{\beta}$ feature from a spectrum
taken on September 8th, 2005, showing a narrow P-Cygni profile.
The red arrow marks the
velocity offset of the blue edge of the absorption feature.}
\label{sn2005db-hbeta-zoom}
\end{figure}

We present our spectroscopy of SN2005db in Fig.~\ref{specfig05db}. 
Early spectra have very smooth continua with composite Balmer 
emission lines clearly showing narrow and intermediate-width 
components as well as narrow P-Cygni absorption features. Weak
He I $5876$\AA\ is also seen in the first two spectra. The October
spectrum shows developing intermediate-width P-Cygni profiles in the Balmer
series, He I, Ca IR triplet and emerging Fe II lines near $5000$\AA. Later 
spectra seem dominated by host galaxy emission, in accordance 
with the light curve (Fig.~\ref{figlc05db}) showing a steep drop in the 
SN luminosity. 
In Fig.~\ref{decomp2005db}, we zoom
in on the $H_{\alpha}$ feature in the August spectrum and extract
the unshocked wind velocity from the decomposition of the $H_{\alpha}$
feature into different Gaussian components. In Fig.~\ref{sn2005db-hbeta-zoom}, 
we zoom in on the $H_{\beta}$ feature in the September spectrum and extract
the unshocked wind velocity from the P-Cygni profile. The P-Cygni
profile in the $H_{\beta}$ feature in the October spectrum was also
analyzed and provides consistent results.

\clearpage

\section{Physical parameters of CCCP SNe IIn}

\subsection{Wind velocity calculation}

We measured the widths of the intermediate and narrow
components of the H$\alpha$ emission line in our SN spectra and calculated 
the corresponding velocities using the Doppler formula. For each event the spectrum where 
the intermediate component was most evident was analyzed.

\begin{figure}
\includegraphics[width=1\textwidth]{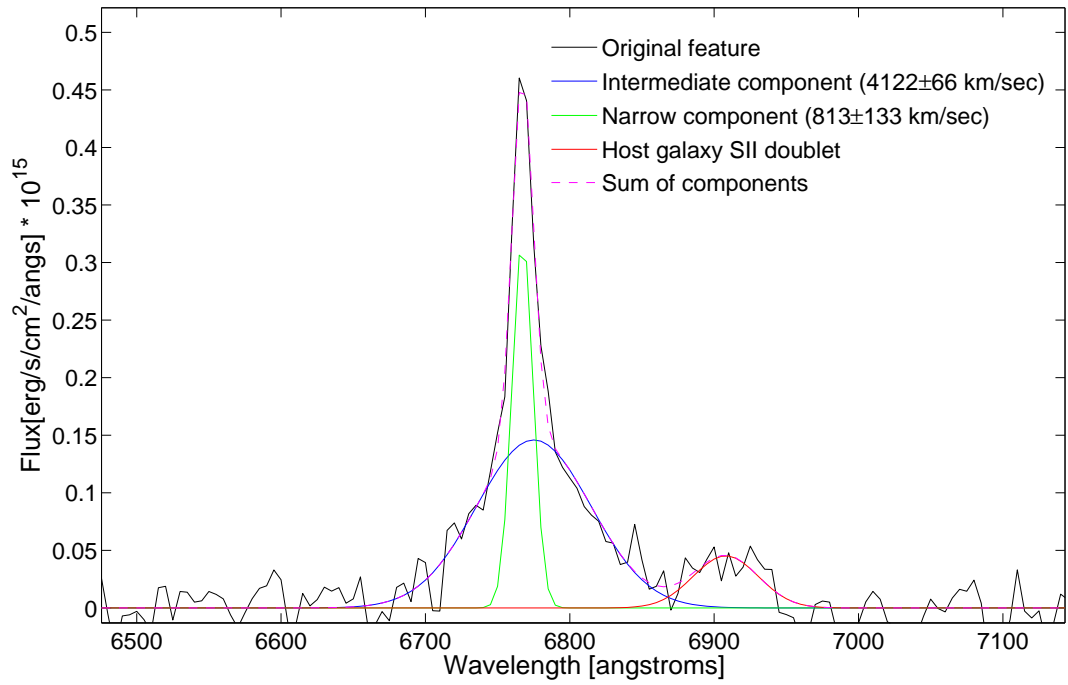}
\caption{The decomposition of the $H_{\alpha}$ feature 
in the 30th of April 2005 spectrum of SN 2005bx.}
\label{decomp2005bx}
\end{figure}

\begin{figure}
\includegraphics[width=1\textwidth]{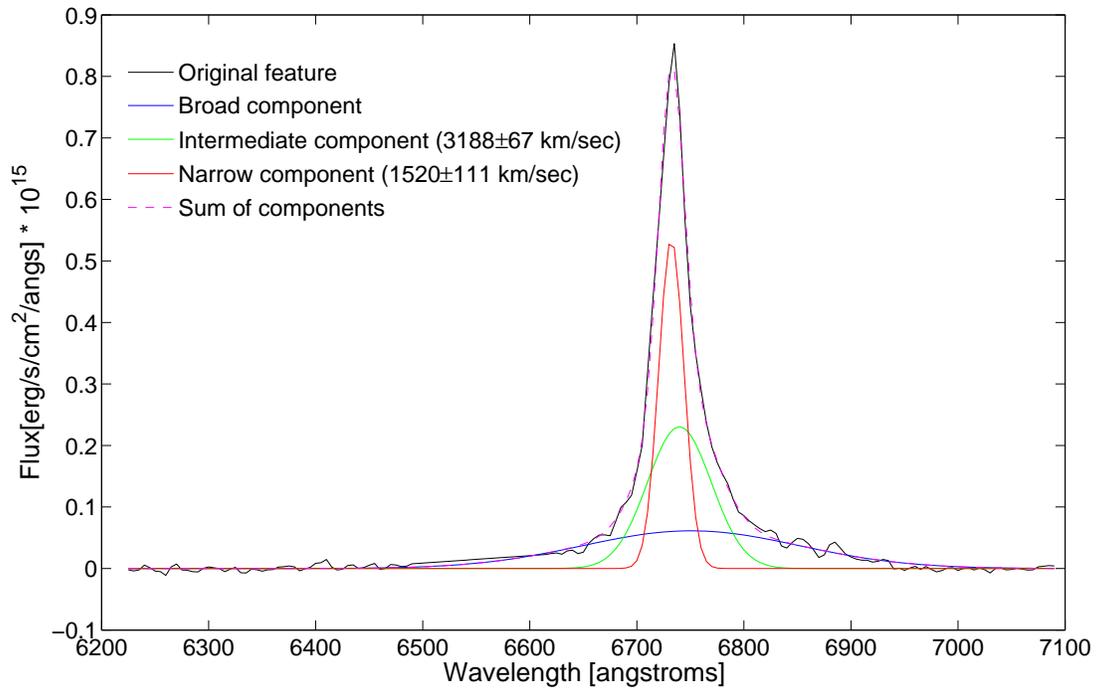}
\caption{The decomposition of the $H_{\alpha}$ feature (post smoothing)
in the 16th of July 2005 spectrum of SN 2005cl.}
\label{decomp2005cl}
\end{figure}

\begin{figure}
\includegraphics[width=1\textwidth]{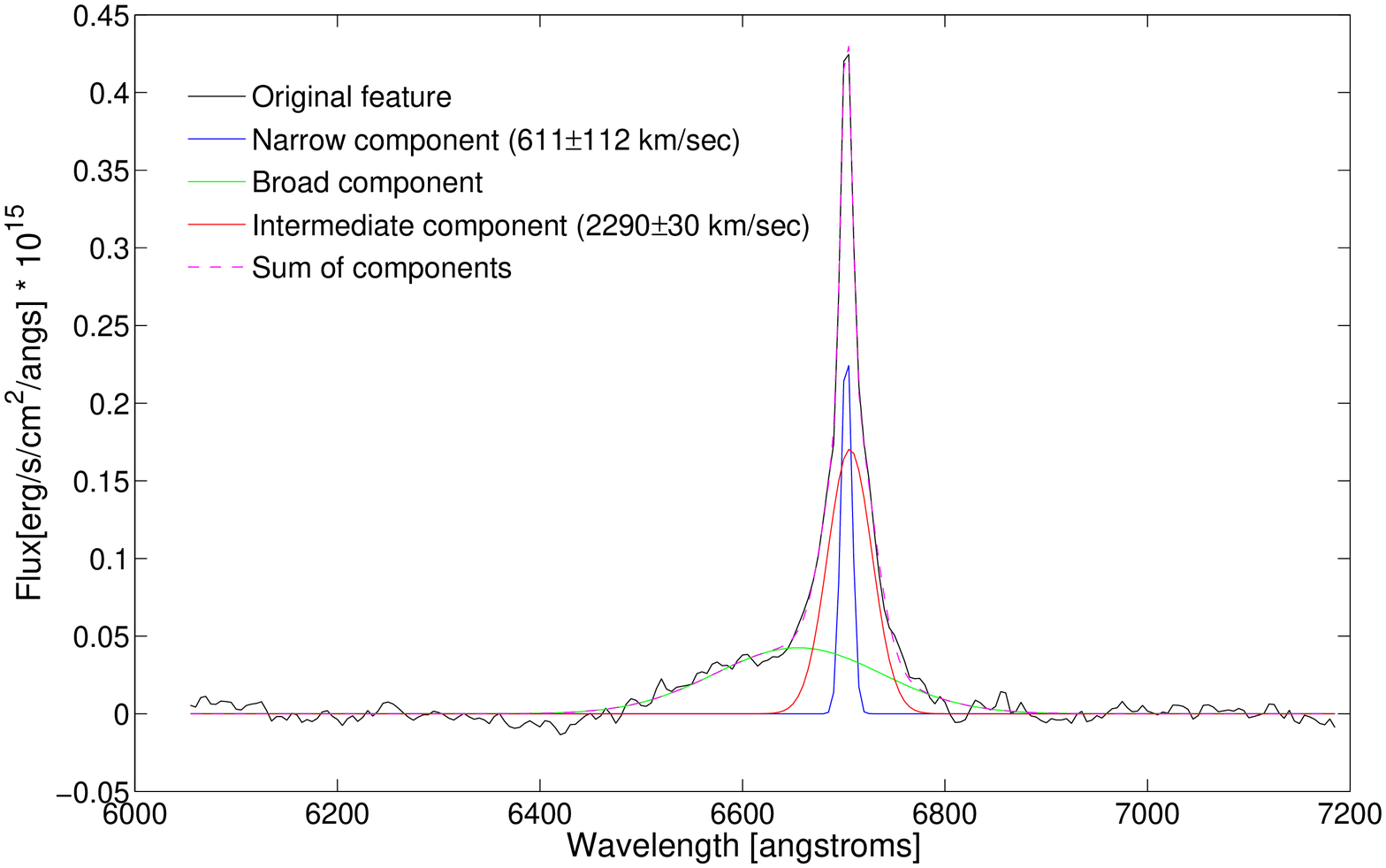}
\caption{The decomposition of the $H_{\alpha}$ feature 
in the 13th of August 2005 spectrum of SN 2005cp.}
\label{decomp2005cp}
\end{figure}

\begin{figure}
\includegraphics[width=1\textwidth]{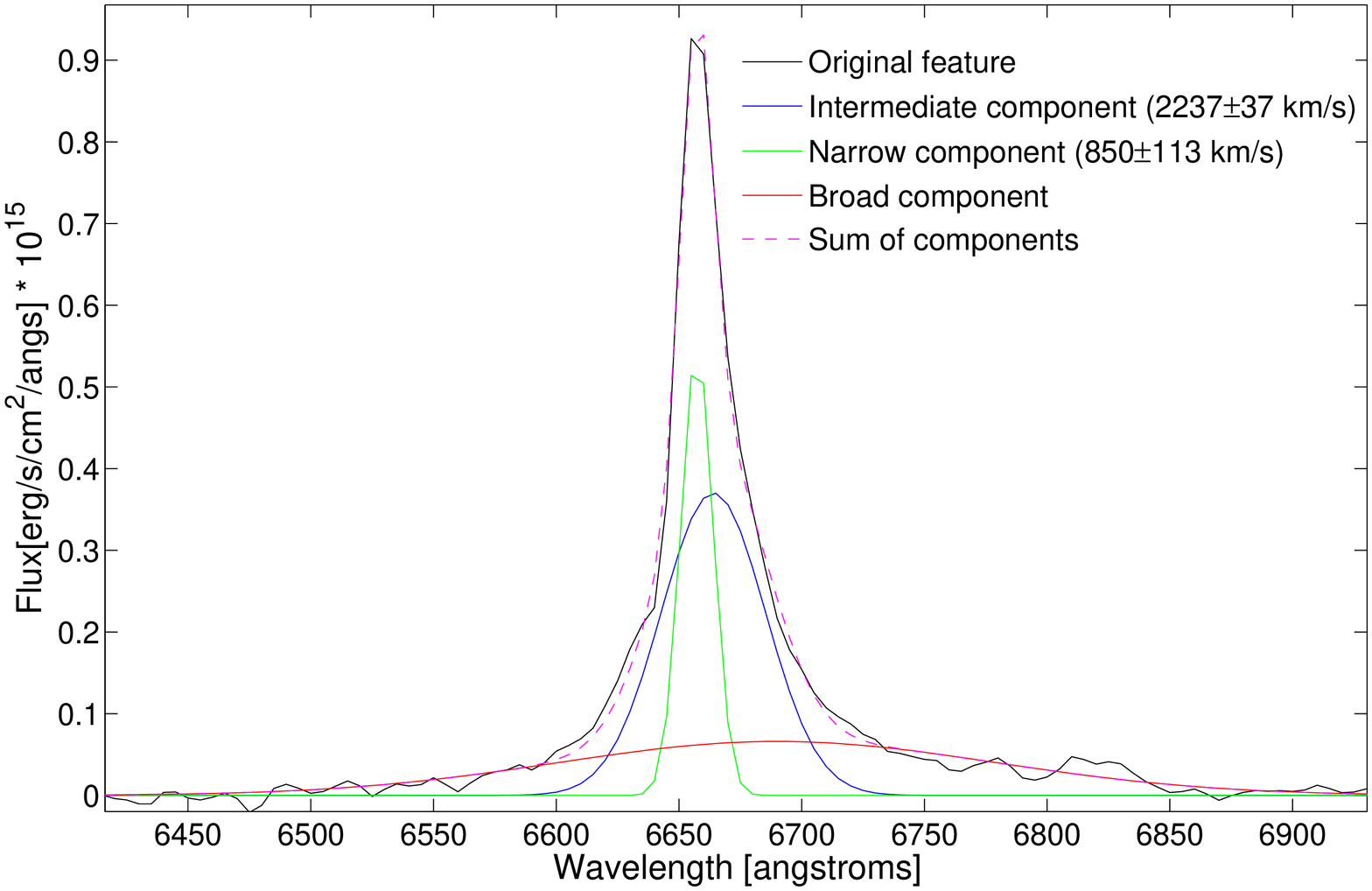}
\caption{The decomposition of the $H_{\alpha}$ feature 
in the 14th of August 2005 spectrum of SN 2005db.}
\label{decomp2005db}
\end{figure}

In order to measure the line widths, the $H_{\alpha}$
feature was decomposed using the following steps, implemented within 
{\it MATLAB}. First, the linear continuum surrounding of the 
line feature was removed. Next, a nonlinear least-squares script solved for the best-fit
multi-Gaussian decomposition. Input parameters consisted
of the number of components and an initial guess for the FWHM, height
and center of each Gaussian, no other constraints were imposed.
The scripts provide the user with the best-fit Gaussian width, height
and center for each component as well as their corresponding errors.
The errors are propagated into the wind velocity error. Line profile
decompositions are shown in Figures~\ref{decomp2005bx} -- \ref{decomp2005db}.
In all cases the best fits required both a narrow and an intermediate-width
component. In some cases a broad component was also required. When
appropriate, narrow P-Cygni profiles were smoothed. The measured 
velocities are reported in Table~\ref{physparamtab}.

\subsection{Mass loss rate calculation}

The calculation is done via the following formula that relates the
mass loss rate of the progenitor with the luminosity resulting from
the ejecta-wind interaction (Chugai \& Danziger 1994):

\begin{equation}
\label{masslosseq}
L_{H_{\alpha}}=\frac{1}{2}\epsilon_{H_{\alpha}}\dot{M}v_{shock}^{2}\frac{v_{shock}}{v_{wind}}
\end{equation}

We assumed that the efficiency factor, $\epsilon_{H_{\alpha}}$ =
0.1, appropriate for young SNe (Salamanca et al. 1998) and compatible
with previous studies. $v_{shock}$ is extracted from the intermediate
component in the $H_{\alpha}$ feature (following the interpretation 
presented in section $\S$~1; but see Chugai 2001; Dessart et al. 2009) 
and $v_{wind}$ is extracted
from the narrow component and/or the blue edge of the narrow P-Cygni
profile (see below).  

To derive the absolute $H_{\alpha}$ flux we performed synthetic photometry 
on the relevant spectra using the methods of Poznanski et al. (2002), compared
the resulting $R$-band magnitude with our photometry (see above, interpolated as
needed) and scaled the spectrum to match the absolute photometry. The
resulting correction factors ranged between 1.7 and 3.2. 
The luminosity was then extracted by integrating over the intermediate-width-feature
in the SN spectrum and accounting for the distance to the SN (obtained from NED).
All measured values are reported in Table~\ref{physparamtab}.

Following previous works, the shock velocity was taken as the
FWHM of the intermediate-width component. The errors
derived from the Gaussian decomposition of the $H_{\alpha}$ profile
are provided by the least squares script output. Running the script
with a different number of components (larger or smaller) turned out
to provide poorer decompositions. 

The unshocked wind velocity
was derived from the HWZI of the narrow component. In
addition to the error calculation procedure described above, 
the errors in the unshocked wind velocity also account for the uncertainty
in the level of the continuum that translates to an uncertainty
in the flux zero point when measuring the HWZI. All of the errors
were propagated linearly into the velocity errors. The unshocked wind
velocity was also calculated via the measurement of the P-Cygni profile,
while taking the width as the difference between the $H_{\alpha}$
or $H_{\beta}$ features peak and the blue edge of
the absorption feature (Fig.~\ref{sn2005cl-halpha-zoom},
~\ref{sn2005cp-halpha-zoom},~\ref{sn2005db-hbeta-zoom}). 
The errors of the P-Cygni velocities
are calculated according our instrumental resolution, and 
the measured values are also 
corrected for the instrumental resolution --
$4.88$\AA/pixel in the red band and $1.08$\AA/pixel in the
blue band, following Alexander, Young \& Hough (1999). These
measurements are also reported in table~\ref{physparamtab}.

For SN 2005cl and SN 2005cp, the unshocked-wind-velocity values used
in the mass-loss-rate calculation include the range
of values consistent with both the Gaussian fitting and the
P-Cygni profile values. The average of the values in this range was
taken as the unshocked wind velocity and the edges of this range determined
the velocity error. For SN2005db, there was no overlap between the
velocities derived by the two methods, so we considered the entire range of values
defined by both methods (from the lower value of the Gaussian
fitting to the top value of the P-Cygni profile). 
The final results can be seen in table~\ref{physparamtab}. 

The mass loss rates derived from Eq.~\ref{masslosseq} using the values derived above
and their respective errors are reported in the final column of table~\ref{physparamtab}.

\begin{table}
\begin{scriptsize}
\begin{tabular}{cccccc}
\hline 
Supernova & Shocked wind & unshocked wind & unshocked wind & unshocked wind  & Mass loss \tabularnewline
          & velocity [km s$^{-1}$] & velocity via Gaussian & velocity via P-Cygni & velocity used for  & rate [M$_{\odot}$ y$^{-1}$]\tabularnewline
          &                         & fitting [km s$^{-1}$] & profile [km s$^{-1}$] & mass loss [km s$^{-1}$] & \tabularnewline
\hline
\hline 
SN2005cl & $3188\pm67$ & $1520\pm111$ & $1318\pm223$ & $1475\pm65$ & $0.12\pm0.02$\tabularnewline
\hline 
SN2005cp & $2290\pm30$ & $611\pm112$ & $632\pm225$ & $610\pm110$ & $0.026\pm0.005$\tabularnewline
\hline 
SN2005db & $2237\pm37$ & $850\pm113$ & $1113\pm65$ & $958\pm220$ & $0.057\pm0.024$\tabularnewline
\hline 
SN2005bx & $4122\pm66$ & $813\pm133$ & N/A & $813\pm133$ & $0.037\pm0.019$\tabularnewline
\hline
\end{tabular}
\end{scriptsize}
\caption{Wind velocities and mass loss rates of SNe 2005bx, 2005cl, 2005cp,
2005db.}
\label{physparamtab}
\end{table}

\section{Previously studied SNe IIn}
\label{historicSNeIIn}

SNe IIn are rare. We have found only 15 SNe IIn for which progenitor
mass-loss rates have been estimated in the literature, and few other
events are discussed in detail (Table \ref{oldSNIIntable}). 
For a single event (SN 2005gl) the progenitor was identified as a Luminous
Blue Variable (LBV)-like star (Gal-Yam et al. 2007; Gal-Yam \& Leonard. 2009)
via pre-explosion images. SNe IIn exhibit a wide range of wind velocities and
mass loss rates (tables~\ref{oldSNIIntable}). 

In addition to conventional SNe IIn, there are other SNe that
exhibit IIn characteristics. One group are the Ia/IIn Hybrids, also
known as type IIa SNe (Deng et al. 2004). 
There is an argument whether these are explosions
of white dwarfs interacting with circumstellar medium (Hamuy et al. 2003; 
Aldering et al. 2006) or the result of a core-collapse in massive stars (Benetti
et al. 2006). 

Another group that technically resembles SNe IIn are faint events that
exhibit very narrow emission lines. These are often understood as
LBV super-eruptions that leave the star intact (so-called ``SN impostors'')
(Van Dyk 2006; Maund et al. 2006). We will not discuss these further here. 

A review of literature events follows.

$\bullet$~SN 1987F is probably the first well-observed SN IIn, and thus
predates the recognition of the class (and its definition) by Schlegel (1990).
Spectroscopic observations were reported by Filippenko (1989) and Wegner \& Swanson
(1996). The event shows both narrow and intermediate-width components. The 
spectroscopic evolution of this object is quite similar to that of SN 2005cp
in our sample (comparing Fig. 2 from Filippenko 1989 with our Fig~\ref{specfig05cp}).
Photometric data have been presented by Filippenko (1989), Cappellaro et al. (1990),
Wegner \& Swanson (1996) and Tsvetkov (1989). 
The object was discovered after peak, but its reported magnitude
($M_{V}=-18.3$\,mag; Filippenko 1989) is similar to that of SN 2005cp (and the average value of
our CCCP sample), while the $V$-band decline rate reported by Cappellaro et al. (1990) 
and Wegner \& Swanson (1996; $\sim0.01$\,mag\,day$^{-1}$)
is also similar to that of SN 2005cp. Chugai (1991) derives a pre-explosion 
mass-loss rate using a formula similar to Eq.~\ref{masslosseq} of $\dot{M}=10^{-3}$\,M$_{\odot}$\,y$^{-1}$,
assuming an unshocked wind velocity of $10$\,km\,s$^{-1}$. Replacing this with the
measured value of the narrow-line width from Wegner \& Swanson (1996; $150$\,km\,s$^{-1}$)
would drive this value up to $\dot{M}\approx10^{-2}$\,M$_{\odot}$. 
 
$\bullet$~SN 1988Z is the first extensively studied event, and is frequently mentioned
in the literature. It was discovered post-maximum, and early photometric and 
spectroscopic observations are presented by Stathakis \& Sadler (1991), 
Turatto et al. (1993a), Filippenko (1997) and Aretxaga et al. (1999). 
Chugai \& Danziger (1994) estimate 
the pre-explosion mass loss rate to be between $7\times10^{-4}$ and $1.5\times10^{-2}$\,M$_{\odot}$\,y$^{-1}$. 

$\bullet$~SN 1994W is a well-studied event discovered before maximum light and extensively
followed. The general properties of this SN closely resemble our observations of SN 2005cl,
with a similar spectral shape, absolute magnitude, rise and decline rates (Sollerman et al. 
1998). These authors measure the unshocked wind velocity from the blue edge of the narrow 
P-Cygni profile to be $1000$\,km\,s$^{-1}$. Chugai et al. (2004) present an extensive analysis
of additional observations, and derive a very high mass-loss rate of 0.3\,M$_{\odot}$\,y$^{-1}$
using similar procedures to those we have employed. Interestingly, Dessart et al. (2009)
have recently suggested that this event may have resulted from a collision between 
consecutive shells ejected by the progenitor, without the final core-collapse SN having
occurred yet. 

$\bullet$~SN 1994aj was studied by Benetti et al. (1998). Its spectra show some similarity
to those of SN 2005cp, but its light curve decays significantly more rapidly 
($\sim0.4$\,mag\,day$^{-1}$), though this decline rate slows down by factor of 10 or so 
at late times. Estimating the unshocked wind velocity to be $\sim1000$\,km\,s$^{-1}$ from
the observed narrow P-Cygni profile and the shock velocity to be around $3700$\,km\,s$^{-1}$
from the HWZI of the H$_{\alpha}$ emission line, these authors derive a mass-loss rate of
$\sim10^{-3}$\,M$_{\odot}$\,y$^{-1}$, using Eq.~\ref{masslosseq}.

$\bullet$~SN 1995G is another well-observed event. Pastorello et al. (2002)
present extensive photometric and spectroscopic observations. Even though the peak was
not observed, the data provided by these authors show this event appears
to be similar to SN 2005cl in terms of its estimated peak magnitude (M$_V=18.5$\,mag)
and its initial decline rate ($\sim0.01$\,mag\,day$^{-1}$). At later times the
decline slows down substantially. The spectra are similar to those of SN 2005cl 
presented above, including the blue continuum ``bump'' which is partially resolved
into individual narrow P-Cygni profiles of numerous lines. Pastorello et al. (2002)
estimate a moderate-mass rate ($\sim2\times10^{-3}$\,M$_{\odot}$\,y$^{-1}$) using 
Eq.~\ref{masslosseq}. However, replacing their assumptions with values consistent
with our methodology (efficiency $\sim0.1$ and wind velocity estimated from the
blue edge of the narrow P-Cygni profile, which would be $\sim1200$\,km\,s$^{-1}$
for this object), would drive the deduced mass-loss rate to very high values
($>1$\,M$_{\odot}$\,y$^{-1}$). Indeed, more sophisticated modeling of the same 
data set by Chugai \& Danziger (2003) explains the properties of SN 1995G with an
explosive mass ejection a few years before the supernova and leads to a mass-loss rate
of $\sim0.1$\,M$_{\odot}$\,y$^{-1}$.

$\bullet$~SN 1995N occurred in a very nearby galaxy (24 Mpc), but was discovered 
long after explosion. Late-time optical and UV observations of SN 1995N are presented 
by Fransson et al. (2002). These authors measure the unshocked wind velocity to be
$<500$\,km\,s$^{-1}$, and estimate the intermediate-width component velocity to
be around $5000$\,km\,s$^{-1}$, but do not estimate the mass loss rate of the progenitor.
Zampieri et al. (2005) use X-ray data to estimate a mass-loss rate of $2\times10^{-4}$\,M$_{\odot}$\,y$^{-1}$.
Earlier observations verbally described by Baird et al. (1998) were apparently 
never published.

$\bullet$~SN 1996L is described by Benetti et al. (1999). Its light curve decays 
rapidly ($\sim0.5$\,mag\,day$^{-1}$), though this decline rate slows down by factor of 5 or so 
at late times. Observations began after maximum light, but these authors estimate a peak
magnitude similar to that of SN 2005cp. 
Estimating the unshocked wind velocity to be $\sim1600$\,km\,s$^{-1}$ from
the observed blue edge of the narrow P-Cygni profile and the average shock velocity 
to be around $4800$\,km\,s$^{-1}$, these authors derive a mass-loss rate of
$\sim10^{-3}$\,M$_{\odot}$\,y$^{-1}$, using Eq.~\ref{masslosseq}.

$\bullet$~SN 1997ab was described by Hagen et al. (1997) and Salamanca
et al. (1998). These authors use high-resolution spectroscopy to accurately
measure the narrow P-Cygni profiles of the Balmer lines, and derive 
velocities of $90$\,km\,s$^{-1}$ for the unshocked wind, and $\sim6600$\,km\,s$^{-1}$
for the shock. Using Eq.~\ref{masslosseq} they derive a mass-loss rate of 
$\sim10^{-2}$\,M$_{\odot}$\,y$^{-1}$. 

$\bullet$~SN 1997eg was studied by Salamanca et al. (2002) spectroscopically. 
Photometric data are not reported. These authors identify a narrow P-Cygni
profile in the Blamer lines, and use its blue edge to determine the unshocked 
wind velocity to be approximately $160$\,km\,s$^{-1}$. Estimating the shock velocity
to be $\sim7000$\,km\,s$^{-1}$, and using the same analysis used for SN 1997ab (and
in out paper), they measure a mass-loss rate of $8.3\times10^{-3}$\,M$_{\odot}$\,y$^{-1}$. 
Tsvetkov \& Pavlyuk (2004) report a few photometric measurements of this event, while
Hoffman et al. (2008) present a model for the star+CSM based on spectrophotometric and spectropolarimetric
observations. 

$\bullet$~SN 1998S is the most extensively observed SN IIn, and probably among
the best-studied SN of any type. Nearby ($\sim$17\,Mpc) and luminous, it was
one of the brightest SNe of the last decades, and has been observed across the
electromagnetic spectrum for many years. Early photometric and spectroscopic
observations are reported by Liu et al. (2000), showing the SN peaked at
M$_B=-18.7$, broadly similar to the objects we studied. Its rise time ($<20$\,days) 
appears short compared to our sample of SNe IIn, and its decay is among the 
fastest ($\sim0.05$\,mag\,day$^{-1}$) measured for SNe IIn. 
Its early spectra resemble those of SN 2005cp (Fig.~\ref{specfig05cp}), in
particular in the H$_{\alpha}$ profile and the broad lines on the blue side
$1-2$\,months after peak. Additional spectroscopy is presented by Anupama et al.
(2001), who derive a mass loss-rate of $10^{-4}$\,M$_{\odot}$\,y$^{-1}$ assuming
an unshocked wind velocity of $10$\,km\,s$^{-1}$; we note that this value
should be corrected upwards by a factor of a few given the wind velocities
estimated by other authors from details analyses (e.g., v$_{w}=50$\,km\,s$^{-1}$,
Leonard et al. 2000; v$_{w}=30$\,km\,s$^{-1}$ Bowen et al. 2000). The analysis
of Lentz et al. (2001) using synthetic spectroscopy modeling indicates
a mass-loss rate of $10^{-3}-10^{-4}$\,M$_{\odot}$\,y$^{-1}$ and wind speeds
of $100-1000$\,km\,s$^{-1}$. Analysis of the radio and X-ray observations
of SN 1998S leads to similar estimates of the mass-loss rate (Pooley et al. 2002).  
Leonard et al. (2000) present an extensive spectroscopic
and spectropolarimetric analysis, while Fassia et al. (2000, 2001) study large
optical/IR photometric and spectroscopic data sets. Optical and UV {\it HST} 
observations of SN 1998S are discussed by Bowen et al. (2000) and 
Fransson et al. (2005). The IR properties of this object are discussed
by Gerardy et al. (2000) and Pozzo et al. (2004), and modeling of the
spectroscopy is further discussed by Chugai (2001) and Chugai et al. (2002). 

$\bullet$~SN 2005gl was studied by Gal-Yam et al. (2007) and Gal-Yam \& Leonard
(2009). This object appears to show a weaker interaction signature than 
other events discussed above, leading to a moderate peak magnitude (M$\sim-17$\,mag)
and relatively fast decay ($\sim0.04$\,mag\,day$^{-1}$; Gal-Yam et al. 2007), as
well a relatively rapid transition from an interaction-dominated spectrum to 
one similar to normal SNe II (Gal-Yam \& Leonard 2009). Analysis by these authors
using methods similar to those used in the current paper lead to estimates of the
unshocked wind velocity v$_{w}=420$\,km\,s$^{-1}$ (confirmed by Smith et al. 2010
using higher-resolution spectra); the shock velocity 
v$_{s}=1500$\,km\,s$^{-1}$) and the mass-loss rate $0.03$\,M$_{\odot}$\,y$^{-1}$.
The analysis of Gal-Yam \& Leonard provides strong evidence that the progenitor of 
this SN was a luminous LBV star.  

$\bullet$~SN 2005ip was studied in detail by Smith et al. (2009). 
This object was not very luminous, with a peak magnitude of 
$\sim-17.4$\,mag, and a relatively rapid decline 
($\sim0.02$\,mag\,day$^{-1}$). The velocity of the unshocked wind is 
estimated to be $100-200$\,km\,s$^{-1}$, and the shock velocity 
(measured from the intermediate-width component) is about $1100$\,km\,s$^{-1}$. 
The mass-loss rate is estimated to be approximately $2\times10^{-4}$\,M$_{\odot}$\,y$^{-1}$.
The spectroscopic analysis supports suggestions concerning dust formation in the ejecta
inferred from IR observations of this object by Fox et al. (2009). 

$\bullet$~SN 2006gy attracted intensive attention due to its very high peak 
luminosity (M$_V\sim-22$\,mag), and it seems some late-time studies may still
be going on, so a complete review of the literature may be premature. Early 
results were reported by Ofek et al. (2007), estimating a mass-loss rate
of $0.1$\,M$_{\odot}$\,y$^{-1}$, and Smith et al. (2007), who were more ambiguous about
such high mass-loss values. Later analysis by Agnoletto et al. (2009), Smith et al.
(2010) and Miller et al. (2010a) do seem to converge on a model invoking an
extremely massive star experiencing strong CSM interaction. Smith et al. (2010)
estimate the CSM velocity to be around $200$\,km\,s$^{-1}$, the shock velocity to be
around $4000$\,km\,s$^{-1}$, and mass-loss rates on the order of $1$\,M$_{\odot}$\,y$^{-1}$.

$\bullet$~SN 2006tf has been studied by Smith et al. (2008). This 
supernova experienced strong interaction with a very massive CSM, making the
event very luminous (the object was caught already in decline, at 
M$_R\approx-20.8$\,mag). Assuming the unshocked CSM velocity at
v$_w=190$\,km\,s$^{-1}$ and the shock velocity at v$_s=2000$\,km\,s$^{-1}$,
these authors estimate very large mass-loss rates $0.1-0.2$\,M$_{\odot}$\,y$^{-1}$
several decades prior to explosion, rising to $2.3-4.1$\,M$_{\odot}$\,y$^{-1}$
a few years before the supernova. 

$\bullet$~SN 2008iy was a remarkable event, with an unprecedented rise time
of $\sim400$\,days, to its peak magnitude of M$_r=-19.1$\,mag (Miller et al. 2010b). 
These authors measure the unshocked wind velocity to be $\sim100$\,km\,s$^{-1}$ (perhaps
with components extending to $450$\,km\,s$^{-1}$), and estimate the
shock velocity to be around $5000$\,km\,s$^{-1}$. They estimate the mass-loss rate
in several ways, getting consistent results of $1 - 2\times10^{-2}$\,M$_{\odot}$\,y$^{-1}$.

$\bullet$~A group of objects whose first well-studied member was SN 2002ic (Hamuy et al. 2003)
have similarities with SNe IIn in that they show strong CSM interaction, but the nature
of the inner explosion (thermonuclear explosion of a white-dwarf star or a core-collapse
of a massive star) are still debated (e.g., Benetti et al. 2006; Trundle et al. 2008). 
Members tentatively included in this group are SN 1997cy (Germany et al. 2000; Turatto et al. 2000),
SN 1999E (Rigon et al. 2003), SN 2002ic (e.g., Hamuy et al. 2003; Kotak et al. 2005) 
and SN 2005gj (Aldering et al. 2005; Trundle et al. 2008). Estimated mass-loss rates 
are within the same range seen for SNe IIn ($10^{-4} - 10^{-2}$\,M$_{\odot}$\,y$^{-1}$; 
Kotak et al. 2004, Trundle et al. 2008).

$\bullet$~In addition to the objects listed above, several additional 
events are mentioned in the literature, on which little information
is provided . These include SN 1978G (Schlegel 1990), SN 1978K 
(discovered long after explosion), SN 1984E (Gaskel 1984; Dopita et al. 1984
Henry \& Branch 1987) SN 1987B (Schlegel 1990, Schlegel et al. 1996), 
SN 1987C (Schlegel 1990; Schlegel \& Kirshner 1998), SN 1988I (Filippenko 1989), 
SN 1989C (Schlegel 1990; Turatto et al. 1993b), SN 1990S (Hamuy et al. 1993),
SN 1994Y (Filippenko 1997; Ho et al. 2001; Tsvetkov \& Pavlyuk 1997); 
SN 1994ak (Filippenko 1997); SN 1999el (Di Carlo et al. 2002);
SN 2007rt (Trundle et al. 2009); SN 2003ma (Rest et al. 2009) and 
SN 2008fz (Drake et al. 2010). Some other recently reported events
whose connection to SNe IIn is unclear are SN 2008es (Gezari et al. 2009;
Miller et al. 2009) and SN 2007od (Andrews et al. 2010).

\begin{table}
\begin{tabular}{ccccc}
\hline 
Supernova & Unshocked wind & Shocked wind   & Mass loss rate         &  References\\
          & velocity       & velocity       & rate                   &            \\
          & [km s$^{-1}$]  & [km s$^{-1}$]  & [M$_{\odot}$ y$^{-1}$] &            \\
\hline
\hline 
SN 1987F  & 150            & 6000           & 10$^{-2}$             & [1] [2] \\
\hline 
SN 1988Z  & $<200$         & 1200 - 1800    & $7\times10^{-2}-1.5\times10^{-2}$ & [3] [4]\\
\hline 
SN 1994W  & 1000           & $\sim$4000     & 0.3                   & [5] [6]\\
\hline 
SN 1994aj & 1000           & $\sim$3700     & 10$^{-3}$             & [7]    \\
\hline 
SN 1995G  & $\sim1000$     & 3000 - 4000    & 0.1                   & [8] [9] [10]\\
\hline 
SN 1995N  & $<500$         & 2500 - 5000    & 2$\times10^{-4}$      & [11] [12]\\
\hline 
SN 1996L  & 1600           & 4800           & 10$^{-3}$             & [13]\\
\hline 
SN 1997ab & 90             & 6600           & 10$^{-2}$             & [14]\\
\hline 
SN 1997eg & 160            & 7000           & 8.3$\times10^{-3}$    & [15]\\
\hline 
SN 1998S  & 30-100         & N/A            & 10$^{-4}$ - 10$^{-3}$ & [16] -- [20]\\
\hline 
SN 2005gl & 420            & 1500           & 0.03                  & [21]\\
\hline 
SN 2005ip & 100 - 200      & 1100           & 2.2$\times10^{-4}$    & [22]\\
\hline 
SN 2006gy & 200            & 4000           & 0.1-1                 & [23] [24]\\
\hline 
SN 2006tf & 190            & 2000           & 0.1-4.1               & [25]\\
\hline
SN 2008iy & 100            & 5000           & 1-2$\times10^{-2}$    & [26]\\
\hline
\end{tabular}

\caption{Wind velocities, mass loss rates and references
for SNe IIn in the literature. References: [1] Wegner \& Swanson 1996; [2] Chugai 1991;
[3] Stathakis \& Sadler 1991; [4] Chugai \& Danziger 1994; [5] Sollerman et al. 1998; 
[6] Chugai et al. 2004; [7] Benetti et al. 1998; [8] Pastorello et al. 2002; [9] this work;
[10] Chugai \& Danziger 2003; [11] Fransson et al. 2002; [12] Zampieri et al. 2005; 
[13] Benetti et al. 1999; 
[14] Salamanca et al. 1998; [15] Salamanca et al. 2002; [16] Anupama et al. 2001; [17] Bowen et al. 2000; 
[18] Leonard et al. 2000; [19] Lentz et al. 2001 ; [20] Pooley et al. 2002; [21] Gal-Yam \& Leonard 2009;
[22] Smith et al. 2009;[23] Ofek et al. 2007; [24] Smith et al. 2010; [25] Smith et al. 2008;
[26] Miller et al. 2010}

\label{oldSNIIntable}
\end{table}

\section{Discussion}

\subsection{Typical properties of SNe IIn}

SNe IIn are rare, and the first few events noticed received some attention as
peculiar type II SN. Following the high-profile work on SN 1988Z and 
the synthesis paper by Schlegel (1990), coining of the term SNe IIn, 
these objects became popular subjects of detailed papers throughout the 
next decade. This trend apparently
died out with SN 1999el, and while the number of detected SNe IIn continued
to rise, no detailed report on a SN IIn discovered between 1999 and 2003 
was published. More recently, several SNe IIn have again been reported, 
but each of these works was motivated by
a conceived peculiarity or special property of its subject event (extreme
luminosity, unusual spectral features, extended light curve, etc.). It thus
seems that while the population of SNe IIn reported in earlier papers
may be fairly representative of the observed population (but often poorly observed
or discovered a long time after explosion), the latest 
papers (typically with better data) are of obviously non-representative events.
It is in this context that the CCCP SNe IIn events are quite useful. Since
the CCCP targeted every core-collapse SN by design, it likely describes more fairly the
SN IIn population detected in nearby galaxies. 

\begin{figure}
\includegraphics[width=1\textwidth]{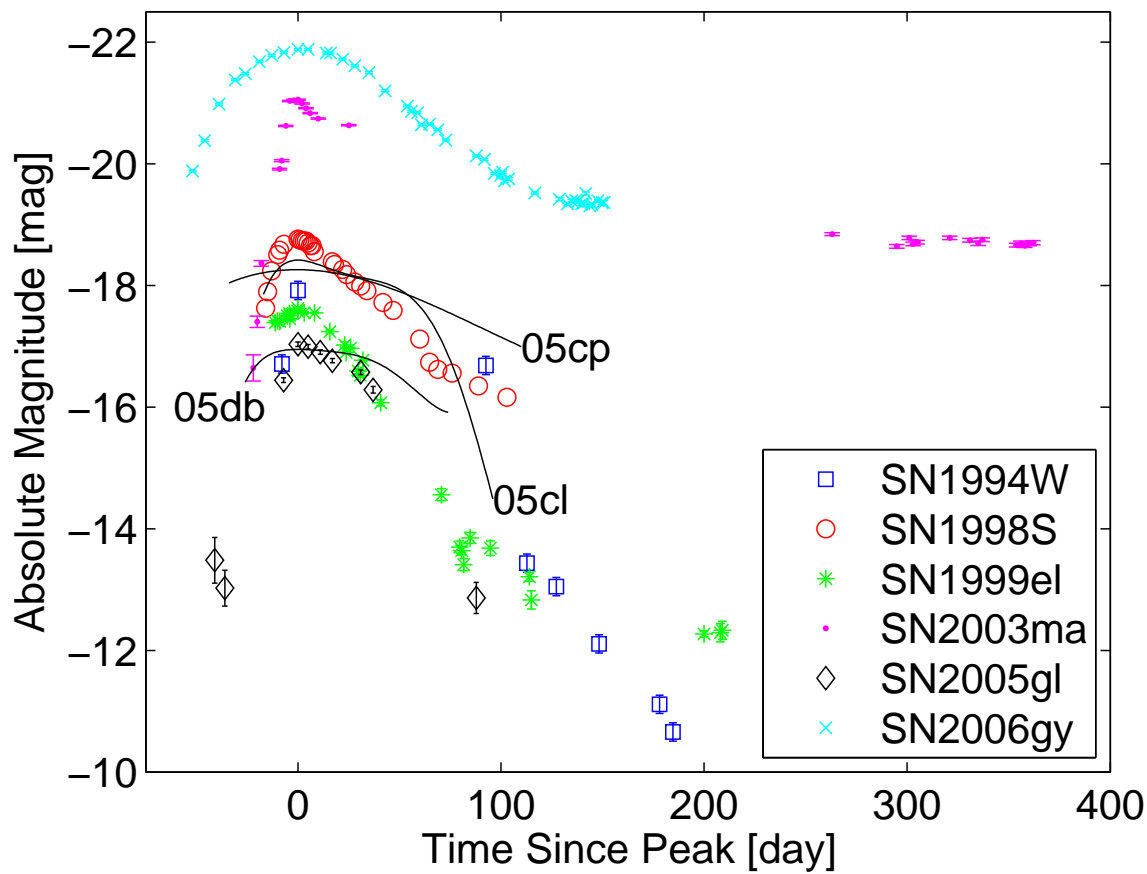}
\caption{$R$-band absolute magnitudes of SNe IIn from the literature and from this 
work. The figure includes every SN IIn for which photometry covering the 
peak is available in the literature (see references in $\S$~\ref{historicSNeIIn}).
The CCCP events are displayed as smooth curves (see~\ref{photsection}) and 
span the range of photometric properties of ``normal'' SNe IIn
in terms of peak magnitude (M$_R=-17 - 18.5$), but appear to have more extended
light curves.}
\label{lit-IIn}
\end{figure}

In Fig~\ref{lit-IIn} we plot the absolute $R$-band light curves of every
SN IIn detected before peak magnitude. As can be seen, the CCCP events
provide a substantial increase to this small sample, and  
span the range of photometric properties in terms of peak magnitude
of what could be viewed as the ``normal'' or ``typical'' population (i.e.,
excluding the recently-reported very luminous events that were singled out
for publication for this reason).   

\begin{figure}
\includegraphics[width=1\textwidth]{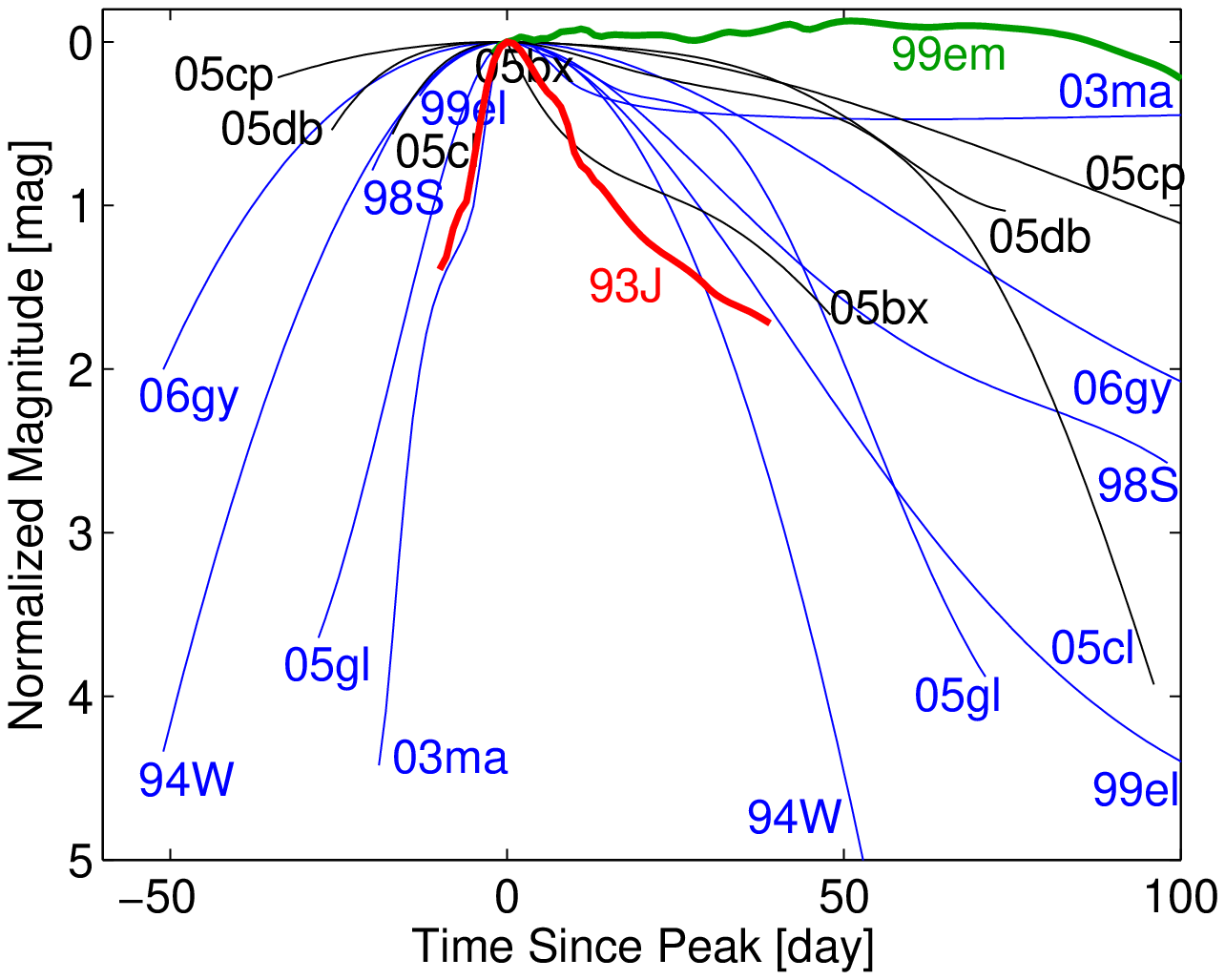}
\caption{The data in Fig.~\ref{lit-IIn} normalized to have the same peak magnitude
and shifted in time so the peaks coincide. We overplot for comparison the light 
curves of the prototypical type II-P SN 1999em, and the rapidly declining type IIb
SN 1993J. One can see that the decline rates of SNe IIn span the range between these
two extremes. On the other hand, the rise times of SNe IIn are typically long, 
extending up to 50 days or more. These data indicate that estimates of explosion
dates and peak magnitudes derived for SNe IIn discovered in the decline phase are
not reliable, as rise times can be long and decline rates vary substantially.}
\label{rise-IIn}
\end{figure}

In Fig.~\ref{rise-IIn} we replot the light curve data normalized and shifted so that
the peaks coincide. Our new CCCP data significantly increase the number of objects
with well-measured rise times, especially if one focused on the ``normal'' luminosity
range. We can see that SNe IIn typically have long rise times, up to 
50 days, and diverse decline rates spanning the range from flat plateau events like
SNe II-P to rapidly decaying events like SNe IIb. Since several events show obvious
breaks or rapid changes in decline rates, the data suggest that extrapolating the
light curves of SNe IIn discovered after explosion backwards in order to determine 
their peak magnitudes may not be a reliable procedure, and that one cannot assume
a negligible rise time when estimating explosion dates from near-peak measurements. 

Spectroscopically, our sample suggests that intermediate-width components and 
narrow P-Cygni profiles in the Balmer lines are ubiquitous among SNe IIn, and
can usually be detected when a time-series of high-quality spectra is examined. 
Interestingly, combining our data with an extensive literature study, it seems
like two spectroscopic groups of SNe IIn seem to manifest. The first, exemplified
in our data by SN 2005cl, displays prominent narrow P-Cygni in the Balmer lines
that persist with time, and a pointed, sharp profile of the H$_{alpha}$ line. 
Other well studied members are SN 1994W and SN 1995G. On the other hand, some 
events, similar to SN 2005cp in our sample, show much weaker narrow P-Cygni
profiles in the Balmer series that may only be visible for short periods of 
time, and have H$_{alpha}$ line profiles that are triangular or boxy at late
times. Other well-studied members include SN 1987F, SN 1994aj and SN 1998S. 
A more complete investigation of the physical reality and nature of this
division lies beyond the scope of this work. 

Finally, it is interesting to note that only one out of the four events 
we studied (with similar data) shows a partially resolved pseudo-continuum in the
blue (Fig.~\ref{pcont}; see Smith et al. 2009 for a recent discussion). 
It is not clear yet if all SNe IIn have blue 
pseudo-continuum dominated by blended Fe-group lines, which is only resolved in a
few cases, or if this phenomenon is rare. It is interesting to note that 
beside being the only event showing a resolved pseudo continuum, SN 2005cl is 
also the SN with the fastest pre-explosion wind velocity and largest mass-loss
value in our sample. A systematic study of a larger sample of SNe IIn 
may also shed light on this aspect. 

\subsection{Implications for the progenitor stars}

The mass loss rates we derive for the CCCP SNe IIn in this work
are indicative of progenitor stars with LBV-like behavior, as 
our values (few $10^{-2}$ to 0.12\,M$_{\odot}$\,y$^{-1}$) are
essentially too high for any class of massive stars other than
LBVs in the eruptive phase (Humphreys \& Davidson 1994). All events
also exhibit high wind velocities, that is some cases 
($1475\pm65$\,km\,s$^{-1}$ for SN 2005cl) are quite extreme 
compared with most known LBVs (though see the recent measurements
by Pastorello et al. 2010 for a counter example). Thus, it is quite possible
that the properties of massive stars shortly before they explode do
not resemble these of stars seen in our galaxy that have not yet 
reached these late, short-lived stages. While the luminous progenitor 
of SN 2005gl (Gal-Yam \& Leonard 2009) and very high deduced mass-loss
rates of some SNe (e.g., SN 2006gy, Smith et al. 2009; SN 2006tf, 
Smith et al. 2008) suggest that some SNe IIn come from extremely massive
stars, it is not clear if this is true for all these events. In this 
context, our measurements of high mass-loss rates argue against the
progenitors of SNe IIn being red giant stars, whose wind velocities 
and mass-loss rates are too low, and the strong hydrogen lines and 
extended light curves argue against Wolf-Rayet stars, whose 
atmospheres are expected to be hydrogen-poor and compact (some
W-R stars may explain the class of interacting H-poor Ibn SNe;
Pastorello et al. 2008). Thus it
seems that SNe IIn as a class (if all or most of the objects arise
from a single group of massive stars) require a progenitor star
that does not fit any of the common classes of massive stars. Such rare
progenitor stars are naturally explained by extreme masses, though other
explanations are possible.

\section{Summary and conclusions}
SNe IIn exhibit a wide range of wind velocities and
mass loss rates (tables~\ref{oldSNIIntable}). 
We have presented our observations of four representative SNe IIn 
observed by the CCCP project. Our photometry allows to determine
peak magnitudes, rise and decline rates for three events. Our
spectroscopic time series are used to measure the pre-explosion
wind velocities and mass-loss rates from the progenitor stars of
these SNe. We then provide an extensive review of the literature
and discuss our results in the context of other well-studied events.
Our main results are:

$\bullet$~SNe IIn are typically luminous compared to other core-collapse
SNe (peak M$_V=-18.4$\,mag), and have a relatively long rise time ($>20$\,days)
followed by a slow decline. 

$\bullet$~SN IIn spectra often show prominent narrow P-Cygni profiles
in the Balmer lines, and multi-component H$_{alpha}$ profiles.

$\bullet$~We measure fast pre-explosion progenitor winds ($600-1400$\,km\,s$^{-1}$)
and derive large mass-loss rates ($0.026-0.12$\,M$_{\odot}$\,y$^{-1}$).

$\bullet$~Our work supports the association between type IIn SNe
and LBV-like progenitor stars. The strong and
massive winds we find are broadly consistent with LBV progenitor stars and unlike
those of typical red supergiants. This is in accordance with the only
well observed IIn SN case where pre-explosion images of the progenitor
were taken (Gal-Yam \& Leonard. 2009). 

\section{Acknowledgments}

We thank N. Chugai, A. Pastorello, S. Benetti and N. Smith for useful comments on the manuscript. The work of A. G. is supported
by grants from the Israeli Science Foundation (ISF), an EU FP7 Marie Curie IRG fellowship,
the Benoziyo Cenetr for Astrophysics and a research grant from the Peter and Patricia
Gruber Awards. S.B.C.~acknowledges generous support from Gary and Cynthia Bengier and the Richard and Rhoda Goldman fund.

\newpage 

\begin{center}
Appendix: photometric calibration
\end{center}

\begin{figure}
\includegraphics[width=1\textwidth]{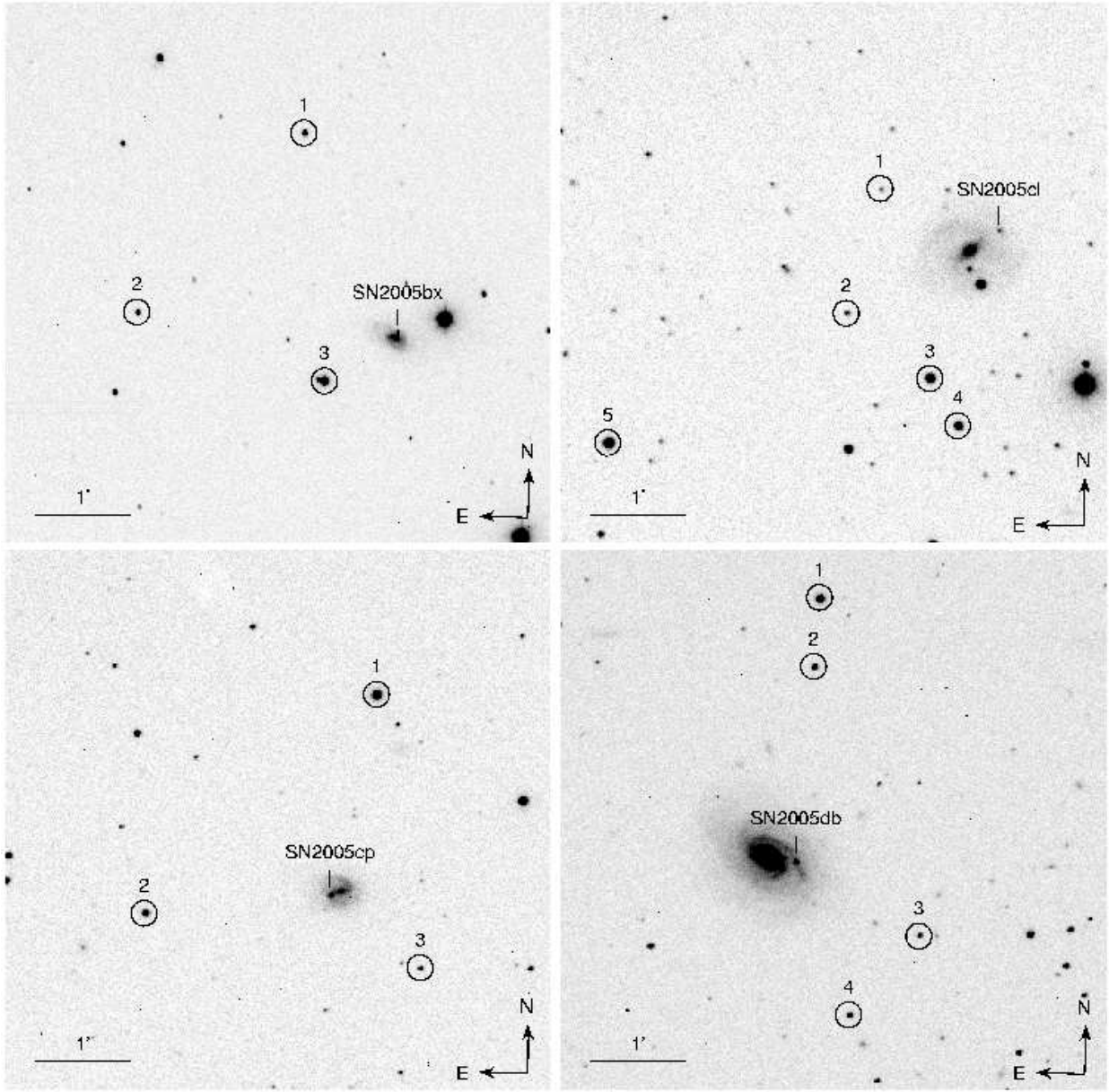}
\caption{Calibration stars in the fields of the SNe studied in this 
paper. The SNe are marked by tick marks, stars are circled and numbered
(matching table~\ref{calstartable}). North is up, East is due left, a
scale bar is provided.}
\label{calcharts}
\end{figure}

In Fig.~\ref{calcharts} we show finding charts for the SNe and nearby calibration
stars, whose magnitudes are given in table~\ref{calstartable}.

\begin{tabular}{llllllll}
\multicolumn{2}{l}{} &  &  &  &  &  & \tabularnewline
\hline
\hline 
 & Star & RA & Dec & B & V & R & I  \tabularnewline
\hline 
SN2005bx & 1 & 13:50:36.5 & +68:35:23.0 & 17.18 & 16.15 & 15.64 & 15.17\tabularnewline
 & 2 & 13:50:54.7 & +68:33:27.0 & 18.06 & 16.72 & 15.99 & 15.35\tabularnewline
 & 3 & 13:50:33.2 & +68:32:48.3 & 14.76 & 14.07 & 13.63 & 13.26\tabularnewline
SN2005cl & 1 & 21:02:07.4 & -06:17:12.5 & 18.54 & 17.97 & 17.65 & 17.26\tabularnewline
 & 2 & 21:02:08.6 & -06:18:30.4 & 18.05 & 17.29 & 16.84 & 16.38\tabularnewline
 & 3 & 21:02:05.0 & -06:19.09.4 & 15.50 & 14.64 & 14.02 & 13.85\tabularnewline
 & 4 & 21:02:03.8 & -06:19:38.2 & 15.91 & 15.12 & 14.58 & 14.34\tabularnewline
 & 5 & 21:02:18.4 & -06:19:57.2 & 14.97 & 14.07 & 13.60 & 13.21\tabularnewline
SN2005cp & 1 & 23:59:44.2 & +18:10:42.2 & 15.03 & 14.28 & 13.83 & 13.44\tabularnewline
 & 2 & 23:59:54.2 & +18:08:23.3 & 17.09 & 16.05 & 15.43 & 14.92\tabularnewline
 & 3 & 23:59:42.1 & +18:07:51.5 & 18.09 & 17.52 & 17.12 & 16.65\tabularnewline
SN2005db & 1 & 00:41:26.0 & +25:32:40.6 & 15.32 & 14.46 & 13.99 & 13.53\tabularnewline
 & 2 & 00:41:26.2 & +25:31:57.4 & 16.44 & 15.87 & 15.56 & 15.22\tabularnewline
 & 3 & 00:41:20.9 & +25:29:12.5 & 18.13 & 17.11 & 16.53 & 15.98\tabularnewline
 & 4 & 00:41:24.1 & +25:28:21.2 & 16.88 & 16.24 & 15.89 & 15.50\tabularnewline
\hline
\hline 
 &  &  &  &  &  &  & \tabularnewline
\label{calstartable}
\end{tabular}

\section{SDSS vs. Landolt calibrations}

Landolt-derived equations for the observing nights of 2005 August 31 and 
2009 April 19 were used to calibrate 15 stars located near 4 CCCP SNe within the
SDSS footprint. The standard Landolt results were compared with magnitudes
obtained from the SDSS photometry of these stars as above. The difference between
the two calibrations was calculated in each filter for each star as
well as the average and the standard deviation of the differences
of all the stars. The results from both nights are shown in table~\ref{crosscaltable}.

\begin{table}[h]
\begin{tabular}{ccccccccc}
\hline 
 &   & 20050831 &   &   &  & 20090419 &  & \tabularnewline
\hline
  & B & V & R & I & B & V & R & I\tabularnewline
\hline 
Average magnitude difference & 0.014 & -0.004 & 0.019 & 0.021 & 0.016 & 0.051 & 0.040 & 0.007\tabularnewline
\hline 
Standard deviation & 0.076 & 0.021 & 0.045 & 0.048 &  0.109 & 0.080 & 0.049 & 0.025\tabularnewline
\hline
\label{crosscaltable}
\end{tabular}
\caption{Comparison between SDSS calibration and Landolt calibrations.}
\end{table}

In all the filters, the scatter in the differences is larger than
the mean offsets. We conclude that the two calibration methods are
consistent. 

\end{document}